%% file: arxiv-template.tex
\documentclass{article}
\usepackage[utf8]{inputenc}
\usepackage{graphicx}
\usepackage{authblk}
\usepackage[a4paper, margin=2cm]{geometry}

\usepackage{amssymb}
\usepackage{latexsym}
\usepackage{eurosym}

\usepackage{url}
\usepackage{xcolor}

\newcommand{\virgol}[1]{``#1''}

\title{Atmospheric muons as an imaging tool}

\author[1]{Lorenzo Bonechi$^{a,}$}
\author[1,2]{Raffaello D'Alessandro$^{b,}$}
\author[3]{Andrea Giammanco$^{c,}$}

\affil[1]{\small Istituto Nazionale di Fisica Nucleare, Sezione di Firenze, Via G. Sansone 1, \newline I-50019 Sesto Fiorentino, Italy}
\affil[2]{\small Universit\`a di Firenze, Dipartimento di Fisica e Astronomia, Via G. Sansone 1, \newline I-50019 Sesto Fiorentino, Italy}
\affil[3]{\small Universit\'e catholique de Louvain, Centre for Cosmology, Particle Physics and Phenomenology, Chemin du Cyclotron 2, B-1348 Louvain-la-Neuve, Belgium}

\begin{document}

\maketitle


\begin{abstract}
\input{abstract.tex}
\end{abstract}

\vspace{9 cm}

{\small $^a$ e-mail: lorenzo.bonechi@fi.infn.it}

{\small $^b$ e-mail: candi@fi.infn.it}

{\small $^c$ e-mail: andrea.giammanco@uclouvain.be (corresponding author)}

\newpage          



\input{introduction.tex}

\input{physics-of-muography.tex}

\input{applications.tex}

\input{detectors.tex}

\input{methods.tex}

\input{issues.tex}

\input{conclusions.tex}

\section*{Acknowledgments}
\input{acknowledgments.tex}

\bibliographystyle{unsrt}
\bibliography{arxiv-template}

\end{document}

%% file: abstract.tex
Imaging methods based on the absorption or scattering of atmospheric muons, collectively named under the neologism ``muography'', exploit the abundant natural flux of muons produced from cosmic-ray interactions in the atmosphere. 
Recent years have seen a steep rise in the development of muography methods in a variety of innovative multidisciplinary approaches to study the interior of natural or man-made structures, establishing synergies between usually disconnected academic disciplines such as particle physics, geology, and archaeology. 
Muography also bears promise of immediate societal impact through geotechnical investigations, nuclear waste surveys, homeland security, and natural hazard monitoring. 
Our aim is to provide an introduction to this vibrant research area, starting from the physical principles at the basis of the methods and reviewing several recent developments in the application of muography methods to specific use cases, without any pretence of exhaustiveness. We then describe the main detector technologies and imaging methods, including their combination with conventional techniques from other disciplines, where appropriate. Finally, we discuss critically some outstanding issues that affect a broad variety of applications, and the current state of the art in addressing them. 



%% file: introduction.tex
\section{Introduction}

The muon ($\mu$) is an elementary particle with quantum numbers in common with the electron but roughly 200 times its mass, abundantly and freely produced in the interaction of primary cosmic rays with the upper atmosphere. 

While opinions diverge on what should be considered the date of discovery of the muon~\cite{Galison1983}, no doubt remained after a crucial experiment about the existence of a new charged particle with mass intermediate between the electron and the proton, published in 1937~\cite{Anderson1937}. 
That was the culmination of years of systematic investigations~\cite{Anderson1961} to understand features of sea-level cosmic radiation that could not fit in the simple theoretical framework of that time, which assumed all elementary charged particles to be protons, electrons, or the recently discovered positron. 
Because of a coincidental agreement in mass, the muon was initially identified with the particle postulated by Yukawa to explain the finite range of the nuclear binding force. 
However, experiments by Conversi, Pancini and Piccioni between 1943 and 1947 demonstrated that the muon has negligible or no nuclear interactions with matter~\cite{ConversiPanciniPiccioni}, in stark contrast with its interpretation as a strong nuclear force mediator. 
Finally, in 1947, Lattes, Muirhead, Occhialini and Powell demonstrated that cosmic muons are produced from the decay of hadrons~\cite{LattesMuirheadOcchialiniPowell}, the most abundant of which (the charged pion, $\pi^\pm$), with mass just slightly larger than the muon, fits the role of the mediator predicted by Yukawa. 
This led to Rabi's famous quip ``Who ordered that?", referred to the muon. 
Today we know twelve fundamental matter particles~\cite{PDG2018}, neatly classified in three generations that only differ by mass, with the muon belonging to the intermediate one; but Rabi's question is still waiting for an answer (its generalized modern version is known as ``the flavour puzzle''~\cite{Feruglio:2015jfa}.)

Several of the features that one century ago helped to clarify this particle's nature are precisely why atmospheric muons are such a good tool for imaging large scale structures (for which, through this document, we employ the neologism ``muography''): absence of strong nuclear interactions, negligible probability of producing electro-magnetic cascades (up to very large momenta, $\approx$~500 GeV~\footnote{To simplify the notation, as customary in particle physics, we drop the $/c$ factor in the momentum units.}), and relatively small energy losses by ionization.  

To the best of our knowledge, the very first practical application of muography dates back to the 1950's, when George studied the feasibility of employing a Geiger counter telescope to infer the ice thickness above a tunnel in an Australian mine~\cite{George1955}. No directional information was available at the detector level, but it could be moved along the tunnel, and the observable of interest was the dependence of the muon flux on the position. 
The first application in archaeology came in the 1960's, by a team led by Alvarez~\cite{Alvarez1970}. 
Their muon telescope was composed of planes of spark chambers and was able to reconstruct particle trajectories, a feature that is taken for granted nowadays but that, as we saw, was missing in George's setup one decade earlier. 

More applications followed. 
A simulation study in 1979 argued the feasibility of muography for mining exploration~\cite{Malmqvist1979}, and compared the traditional telescope geometry (appropriate for easily accessible tunnels) with a set-up that could fit in narrow boreholes.
Several other geophysical applications have been explored, including the imaging of mountains (with the first example in 1995 by Nagamine's team~\cite{Nagamine1995}), which since the beginning had been motivated by future applications to volcanoes. Probably also because of the immediate societal relevance of volcano monitoring in Japan, Japanese teams have been at the forefront of muography since the 1990's. 
A major breakthrough was the first application of muography to predict the eruption sequence of Mount Asama, Japan, during its 2009 unrest~\cite{Tanaka2009}. 
Nowadays, volcanology is probably muography's most developed area of investigation. The list of volcanoes already actively studied includes Asama, Iwodake and Sakurajima in Japan, Vesuvius, Etna and Stromboli in Italy, as well as Puy de D\^ome and La Soufri\`ere de Guadeloupe in France. 
While all the applications described thus far are based on the attenuation of the muon flux in traversing the target, an important breakthrough was the first exploitation in 2003, by Borozdin et al.~\cite{LosAlamos2003}, of the large-angle scattering of the muons due to the intense Coulomb field of nuclei, which allows to discriminate materials based on their atomic number and opens applications in the nuclear sector and in homeland security. 
Muography is nowadays a booming research area, with a steadily growing trend of publications since the beginning of this century and more than 25 articles per year in the last few years\footnote{As reported in Fig.~1 of Ref.~\cite{Kaiser2018}, with data from the University of Glasgow Library. It can be argued that those numbers are most likely on the conservative side, given the amount of workshops, seminars, and conferences that only publish the slides of the various interventions.}.

Several excellent reviews are already available on this subject.  Procureur~\cite{Procureur2018} pays some attention to detector considerations, while Kaiser~\cite{Kaiser2018} includes an interesting discussion on commercial aspects. 
Carloganu and Saracino~\cite{CarloganuSaracino2012} and Tanaka~\cite{Tanaka2017} elaborate on volcano studies and in particular, respectively, on the lessons learned in Europe and Japan. Checchia~\cite{Checchia2016} specifically focuses on scattering-based muography. 
The present article aims at providing a fresh perspective on recent trends, elaborating in particular on the challenges to be overcome in order to make the transition, in all its diverse areas of application, from ``proof of principle'' to an established imaging tool. 

This article is structured as follows: 
Section~\ref{sec:physics} aims at a pedagogical summary of a few physical properties of atmospheric muons that are at the basis of muography. 
Section~\ref{sec:applications} provides a snapshot of the current trends in muography via a non-exhaustive list of recent applications. 
Section~\ref{sec:detectors} outlines the main detector design choices and elaborates on their rationale. 
Section~\ref{sec:methods} presents methodological considerations related to image reconstruction.
Section~\ref{sec:issues} addresses some of the general issues affecting muography research, such as background reduction and estimation, momentum bias, and accuracy/speed trade-off in simulations. 
Finally, we conclude in Section~\ref{sec:conclusions} by summing up the lessons learned, and providing our personal outlook of this vibrant area of research and development.

%% file: physics-of-muography.tex
\section{The Physics of Muography}
\label{sec:physics}

The muons that we exploit for imaging are called ``atmospheric'' because they originate within the atmosphere of our planet (typically 15~km above the sea level~\cite{PDG2018}), or ``cosmic'' because of their origin. 
The primary cosmic rays entering the atmosphere (mostly protons, with small fractions of heavier nuclei, electrons, positrons and antiprotons) collide with the atmospheric nuclei (mostly Nitrogen and Oxygen), producing hadrons that in turn can further interact with the atmosphere, unless they decay. Pions ($\pi^+$, $\pi^0$, $\pi^-$) and kaons ($K^+$, $K^0_S$, $K^0_L$, $K^-$) are abundantly produced in these collisions, and when they are charged their dominant decay modes produce muons~\cite{PDG2018}: $\pi^\pm\to\mu^\pm\nu_{\mu}(\bar\nu_{\mu})$ ($\approx 99.99\%$ of the times) and $K^\pm\to\mu^\pm\nu_{\mu}(\bar\nu_{\mu})$ ($\approx 64\%$ of the times, with a further 3\% where a $\pi^0$ is also produced); when not decaying directly into muons, charged kaons mostly decay into two or three pions ($\pi^\pm\pi^0$, $\pi^\pm\pi^\pm\pi^\mp$ or $\pi^\pm\pi^0\pi^0$), which in turn decay into muons if they are charged. Also the long-lived neutral kaons produce muons $\approx 27\%$ of the times via the decay channel $K^0_L\to \pi^\pm\mu^\mp \bar\nu_{\mu} (\nu_{\mu})$.

Muons are unstable, decaying into electrons and neutrinos ($\mu^+\to e^+\bar\nu_{\mu}\nu_e$ and $\mu^-\to e^-\nu_{\mu}\bar\nu_e$) with a lifetime $\tau \approx 2~\mu {\rm s}$ at rest, but a well-known relativistic effect dilates their observed lifetime by the Lorentz factor $\gamma \equiv 1/\sqrt{1-(v/c)^2} = \sqrt{1+(p/m_{\mu}c)^2}$. Due to the relative hardness of the atmospheric muon spectrum, most of them reach sea level: for example, a momentum of 4~GeV (about the peak value of the muon momentum distribution) corresponds to $\gamma \approx 20$, which naively yields\footnote{A more realistic calculation of $l$, out of the scope of this discussion, should consider the continuous slow-down of the muons while they lose an energy of up to 2~GeV by ionization (see discussion below and Eq.~\ref{eq:dedx}), and the non-homogeneous density of the atmosphere. For $p = 2.4~{\rm GeV}/c$, simulations yield $l\approx 8.4~{\rm km}$~\cite{PDG2018} while our naive calculation yields 15~km. The larger the momentum, the better this approximation works.} a decay length $l = \gamma\beta c\tau \approx 24~{\rm km}$. Most of the Earth's atmosphere is contained within 16~km, and as said above most atmospheric muons are produced at a height of around 15~km.
At sea level, muons constitute the vast majority of the charged particles, arriving at a rate of roughly 100~Hz/m$^2$. 
Most of the other charged particles produced in the cosmic shower, that are able to reach the sea level are protons, electrons and positrons; although they do not decay, these background particles disappear or lose energy much more quickly than muons because of their stronger interactions with matter: nuclear interactions affect protons but are absent for muons; energy loss by bremsstrahlung depends on $1/m^2$ and is therefore 40,000 times smaller for muons than for electrons and positrons; moreover, positrons easily annihilate with the electrons of the atmospheric atoms. 
For these reasons, the combined background flux is much smaller than the muon flux. Above 1~GeV, the muon flux at sea level is at least two orders of magnitude larger than any other charged particle~\cite{PDG2018}, while the contamination of $e^\pm$ grows significantly when the momentum cut-off of the apparatus is lower~\cite{Bonechi2005,Bogdanova2006,SatoKinWatanabe2017,Olah:2017zgo}.

The angular distribution of the atmospheric muon intensity is found to be approximately proportional to $\cos^n \theta$, where $\theta$ is the zenith angle and $n \approx 2$~\cite{Lin2010}, with mild dependence of $n$ on energy, latitude, altitude and depth~\cite{Grieder2001}. 
This has the practical implication that much longer exposure times are needed when it is necessary to use muons at large zenith angles (which is typically the case when studying mountains and volcanoes from a distance) with respect to cases where the large flux of vertical muons can be exploited\footnote{For example, the difference in flux and therefore exposure time between vertical and horizontal orientation was found to be around a factor of eight with the detector geometry of Ref.~\cite{LosAlamos2014}. In general, this ratio depends on the angular acceptance of the detector, over which the angular distribution of the incoming muons is integrated.}.


Ionization and atomic excitation are the dominant energy loss mechanisms for muons in matter up to momenta around 500~GeV. 
The momentum spectrum of atmospheric muons peaks at around 4~GeV; at those momenta, their mean energy loss rate ($-\frac{dE}{dx}$) is close to the minimum of the Bethe function, that in the regime of relevance (and neglecting some small corrections) we can write as:
\begin{equation}
\label{eq:dedx}
    -\frac{dE}{dx} \propto \frac{Z}{A}\cdot \rho \cdot \frac{1}{\beta^2}\cdot\left[ \ln \left( K\cdot \frac{\beta^2}{1-\beta^2} \right) + corrections \right] \, ,
\end{equation}
where $Z$ and $A$ are the atomic and mass number, respectively, of the element characterising the traversed material\footnote{In mixtures and compounds, the average energy loss rate can be approximated by a weighted average.} and $\rho$ is its density, while $\beta = v/c$ is the relativistic speed of the muon, and $K=2m_e c^2/I^2$ where $I$ is the mean excitation potential, of order electronvolts, and $m_e$ is the electron mass ($m_e c^2 = $~511~keV). 
In the approximate formula of Eq.~\ref{eq:dedx}, the dependence on the traversed material comes through $\rho$, $Z$, $A$ and $K$. 
It is important to remark that for intermediate-mass elements, and in particular for those that compose most of the mass of the Earth's crust, the $Z/A$ ratio deviates little from a 1/2. The constant $K$ depends on $I$ which in turn has an approximately linear dependence on $Z$~\cite{PDG2018}; the fact that it only appears logarithmically in the formula mitigates the practical importance of its variation. 
In conclusion, with the only exception of hydrogen, the relationship between the energy loss rate and $\rho$ can be considered quasi independent of other properties of the traversed material. 

Thus, the range of a muon is obtained by inverting and integrating Eq.~\ref{eq:dedx} from the energy at the entry point (where $\beta\approx 1$) up to where the muon is at rest ($\beta = 0$). 
In practice, what is directly measured by a muon detector is the muon flux coming from the various directions within its acceptance; compared to the ``free-sky'' flux, this yields the probability for a muon to be absorbed by a given target along a certain line of sight (also known as ``muon transmission''), which is the basis for \emph{absorption-based muography} (\textbf{AM}), also quoted as \emph{transmission muography}. 
This can be directly interpreted in terms of ``opacity'' along that line of sight, defined as the density integrated along a path length: $O = \int \rho(x) dx$. 
A convenient unit for opacity is \virgol{meters water equivalent} (\textbf{mwe}), with a conversion factor $\rm 1~mwe = 100~g/cm^2$; the energy loss of an energetic muon is about $\rm 0.2~GeV/mwe$. 

In cases where the thickness along the lines of sight of the detector is known, e.g. when imaging a target whose surface is precisely known, the opacity along a direction can be turned into a measurement of the average density along that direction. Conversely, some applications actually have the opposite goal, that of measuring the boundaries of a volume (e.g. the overburden of a tunnel), in this case a fair guess of the average density must be used as input. 
Due to the presence of $\beta$ in Eq.~\ref{eq:dedx}, the AM method also needs as input the momentum spectrum of the atmospheric muons.

As shown by Rutherford in 1911~\cite{Rutherford1911} using very thin targets, charged particles traversing matter are deflected by Coulomb scattering due to the intense electric fields near nuclei. 
The probability distribution of the angular deflection $\Delta\theta$ for a single scattering follows Rutherford's Law, $P(\Delta\theta) \propto 1/\sin^4{(\Delta\theta/2})$. Most deflections are small, so after traversing a macroscopic amount of material the actual distribution features an approximately Gaussian core, as expected from the central limit theorem, asymptotically reaching Rutherford's Law in the tails of $P(\Delta\theta)$ due to the rare large-scattering events. 
The Gaussian approximation gives a good description for 98\% of the actual distribution~\cite{PDG2018}, with a standard deviation that in a broad range of $Z$ and for not-very-thin targets is well approximated by~\cite{LynchDahl1991}:
\begin{equation}
    \label{eq:scattering}
    \sigma(\Delta\theta) = \frac{\rm 13.6~MeV}{\beta \cdot c \cdot p}\cdot \sqrt{\frac{x}{X_0}} \cdot \left[ 1+0.038 \cdot \ln{\frac{x}{X_0\cdot\beta}} \right] \, ,
\end{equation}
where $p$ is the muon momentum, $x$ is the path length from entry to exit, and $X_0$ is the radiation length. 
Very accurate estimations of $1/X_0$ exist~\cite{Tsai1974}, but for sake of illustration we present here a particularly compact formula (accurate to better than 2.5\% for all elements except helium~\cite{GlueX}):
\begin{equation}
    \label{eq:x0}
    \frac{1}{X_0} \propto \frac{Z(Z+1)}{A}\cdot \ln \left( \frac{287}{\sqrt{Z}} \right) \, .
\end{equation}
It is apparent from Eqs.~\ref{eq:scattering} and~\ref{eq:x0} that the root mean square (RMS) of the deflection angle is directly related to $Z$; this is the basis for \emph{scattering-based muography} (\textbf{SM}), which has been exploited since 2003~\cite{LosAlamos2003} for applications where contrast is sought between a high-$Z$ material and a lower-$Z$ background (e.g., fissile nuclear material hidden within scrap metal, or heavy metal within concrete). 
In a 10~cm thick layer, the RMS scattering angle for a 3~GeV muon is 2.3 milliradians in water, 11 milliradians in iron, and 20 milliradians in lead~\cite{LosAlamos2003}. 

The average muon deflection $\langle \Delta\theta \rangle$ due to Coulomb scattering is always zero, independently of the type and amount of material traversed, which only affects the width of its distribution. 
Reference~\cite{Zenoni:2014kva} proposes to exploit this feature for the long-term monitoring of the stability of a building. 
The method requires two muon telescopes, positioned respectively on a structural element of the building (the fixed reference system) and on the point of the building  to be monitored. Any deviation in the apparent $\langle \Delta\theta \rangle$ appearing with time would indicate a deformation of the structure.

Both types of muography, AM and SM, rely on the statistical distributions of large samples of muons (e.g., one cannot deduce the presence of a high-$Z$ material from the trajectory of a single muon). 
Reference~\cite{Procureur2018} illustrates the relative merits of the absorption and scattering methods by listing the survival probability and RMS scattering angle $\sigma(\Delta\theta)$ for multi-GeV muons when traversing a few hypothetical targets. One of the examples concerns the muographic image obtained for the same $10\times 10\times 5~{\rm cm^3}$ lead brick using the two methods with different exposure times, and it shows that for this type of high-$Z$ target, SM needs an order of magnitude less time than AM to achieve a recognizable image, and that it yields an image with better definition. 
As a general rule, AM is a powerful method for very large targets, while SM is more appropriate for small and medium sized targets.


%% file: applications.tex
\section{Applications}
\label{sec:applications}

In this section, we review recent developments in the application of muography methods in geosciences, archaeology, civil engineering, nuclear safety, and security, without pretense of exhaustiveness but with the intention of giving a comprehensive overview of the field, setting the stage for the main methodologies presented in Sections~\ref{sec:detectors} and~\ref{sec:methods}.

\subsection{Geosciences}
\label{sec:geophysics}

The typical geophysical applications of muography feature very large targets, for which AM is the most appropriate technique. 
Muography complements traditional survey methods by offering directional sensitivity to the in-depth structure of large objects. Unfortunately, some of the most pressing questions in geology require that narrow structures in the deepest part of a O(km) target be resolved, even at low elevation angles where the muon flux is less intense. 
In these conditions, not only the signal is faint, but it can also be overwhelmed by backgrounds that can limit the measurements to an insurmountable ``systematics wall'' (i.e., uncertainties that do not decrease with more data). 

Within this context, one of the most widespread applications of AM is the imaging of volcanoes and several partnerships with volcanologists have been established by the physicists involved. 
Different volcanoes have attracted the attention of the muography community for their intrinsic volcanological interest and/or for their natural hazard potential. As this technique is still very young, the choice of the target is often driven at least in part by reasons of opportunity, e.g. geographical proximity to both a strong particle physics laboratory and a strong volcanology institute. This is often the case in countries where both communities are well developed, explaining the number of activities in Japan~\cite{Tanaka2018}, Italy~\cite{Raffaello2018}, and France~\cite{CarloganuSaracino2012,Marteau:2015pxa}. 

\begin{figure}[htbp]
\centering
\includegraphics[scale=.5]{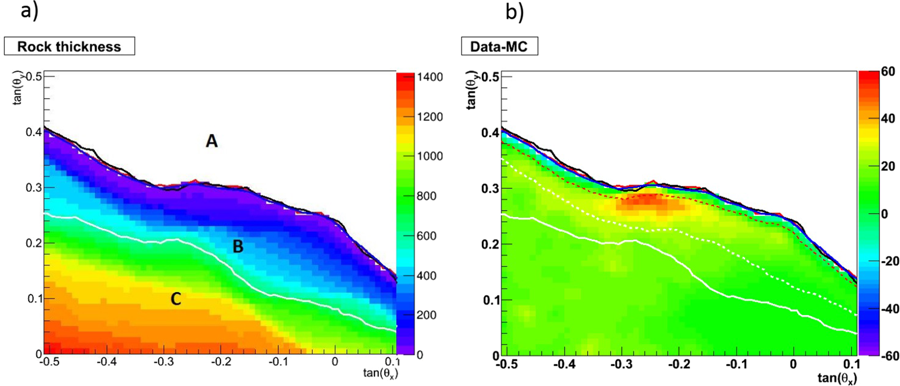}
\caption{(a) Muography of the crater region of Stromboli volcano as seen by an emulsion detector. The color scale is the rock thickness in meters. (b) Difference between the observed muon flux and the one expected from Monte Carlo simulation. Color scale represent muons counts. Reproduced from~\cite{Tioukov2019}.}
\label{fig:stromboli}
\end{figure}

Static pictures using muon data integrated over several months can be used to investigate the inner composition of a volcano providing information on the location of volumes of rocks with different densities, that is essential to the understanding of the volcano's history. 
A recent example is the first muographic imaging of Mt. Stromboli~\cite{Tioukov2019}, a strato-volcano of the Aeolian archipelago (Italy) with a height of about 920~m above sea level, characterized by the emission of huge amounts of gas and a continuous ongoing eruptive activity named ``Strombolian'' in its honour.
The measurement was carried out with emulsion detectors, particularly appropriate to the logistic challenges~\footnote{The access to the optimal observation point is so impervious that the detector had to be pre-assembled at a lower altitude and then lifted up to the observation point by helicopter.} for the reasons explained in Sec.~\ref{sec:emulsions}, taking data for five months. This study found a significant low-density zone ($30-40\%$ contrast with respect to bedrock, see Fig.~\ref{fig:stromboli}) at the summit of the volcano. This result is relevant for geophysics and hazard estimation, as the structural setting of this part of the volcanic edifice controls the eruptive dynamics and the stability of the ``Sciara del Fuoco'' slope, which is affected by recurrent tsunamigenic landslides. 

Hazard concerns are also a major motivation for the study of Mt. Vesuvius, a strato-volcano near Naples (Italy), world-famous for its catastrophic eruption in 79 AD that obliterated the ancient Roman cities of Herculaneum and Pompeii and was vividly described by Pliny the Younger (hence the name ``Plinian'' to indicate one of the major eruptive classes). Nowadays more than half million people reside in the ``red zone'' surrounding Vesuvius, defined as being at high risk of pyroclastic fallout in case of a new Sub-Plinian eruption~\cite{Barberi2013}. 
Mt. Vesuvius' first 2D muographic images were produced by the MU-RAY project~\cite{MURAY2014} using three $x-y$ layers of scintillator bars. 
To reduce backgrounds, the MURAVES project~\cite{Raffaello2018,Saracino:2017mao}, MU-RAY's successor, added a fourth layer and a 60~cm lead absorber between the third and fourth layer, complemented by time-of-flight (TOF) measurement (see Sec.~\ref{sec:bkg}). MURAVES aims to reach sufficient precision to resolve internal discontinuities of about 10~m in the upper part of Mt. Vesuvius. 


The Puy de D\^ome, a dormant volcano near Clermont-Ferrand (France), is a lava dome 1465~m high that is part of a long volcanic chain. 
It constitutes an excellent ``standard candle'' to test muography detectors and methods thanks to the abundance of reference data from standard geophysical methods, and also for the relatively accessible logistic support. 
This opportunity has been exploited for the development of the TOMUVOL telescope~\cite{Carloganu:2011zz,Bene:2013fwa,Carloganu2013}, composed of planes of glass RPC detectors (see Section~\ref{sec:gas}) originally developed in the context of the particle physics CALICE collaboration (R\&D for a detector to be operated at future high-energy linear colliders). 
A joint measurement campaign of the Puy de D\^ome in 2013~\cite{Ambrosino:2015yqk} by the TOMUVOL and MU-RAY collaborations with their detector prototypes, based on two different detector technologies with complementary merits,
was crucial to quantify the challenges and prioritize their further developments. 
They found that, with their detector setups at that time, the backgrounds overwhelmed the signal for opacities larger than 500 mwe from 1~km away; the further characterization of these backgrounds has led to the design of improved detectors with higher momentum threshold and TOF measurement~\cite{Raffaello2018}, with the goal of a robust muographic imaging of kilometer-scale volcanoes. 


Volcanoes are very dynamic systems, therefore nuclear emulsions, in spite of their superior resolution and logistic advantages, are less popular than detector technologies that allow to time-stamp the observed muons. 
An obvious use case is hazard prevention: a temporal evolution in the muon flux through the core of a volcano may be potentially listed as an eruption precursor. 
The data from a muography campaign on Mt. Asama during its 2009 unrest~\cite{Tanaka2009}, showing a temporal variation of the observed muon flux through the crater region that correlated with magma ascensions and descents, were reported to the Japan Meteorological Agency, which used them as one of the inputs for eruption forecast purposes~\cite{JapanMeteorologicalAgency}. 
Moreover volcanoes can have rich hydrothermal systems, affecting the density distributions close to their surfaces.
This is another topic of relevance for hazard prevention, as hydrothermal fields of moderately active volcanoes have an unpredictable behavior; hazardous events that can develop rapidly, with no known precursory signal that is clearly identified as a potential warning of imminent danger. 
Muographic time series have been used in the study of the hydrothermal system of La Soufri\`ere de Guadeloupe~\cite{Jourde2016}, an active volcano in the Lesser Antilles (France) that has been extensively studied by the DIAPHANE collaboration~\cite{Marteau:2016jcn} taking data with several identical muon telescopes based on scintillators from different observation points. In a recent study~\cite{Gonidec:2018kij}, combining muography with seismic noise monitoring, they were able to detect with an unprecedented space and time resolution the increase of activity of a hydrothermal spot located 50 to 100~m below the summit, at timescales of few hours to few days.

Mount Etna, near Catania (Italy), is one of the tallest and most active volcanoes in the world, with a height of more than 3~km and a basal diameter of 40~km. Its eruptions occur through one of its four summit craters or from vents or fissures on its flanks. 
In spite of being one of the most studied volcanoes in the world, the geometry of the shallow conduit network feeding its four summit craters is still largely unknown~\cite{Carbone2013Etna}. 
Etna's size makes it a very challenging target for muography, but a few teams have investigated the feasibility of the method with various approaches and targeting different craters. The first imaging was performed in 2010~\cite{Carbone2013Etna}, targeting the South-East Crater, 240\,m high with a base diameter of 500~m, using an early version of the DIAPHANE telescope. 
Recently, the MEV (Muography of Etna Volcano) project reported preliminary results~\cite{MEV2018} with their dedicated high-resolution telescope, also based on scintillators. In this first study, the extinct Monti Rossi crater was imaged to be used as a reference in view of future studies of the active North-East Crater. 
The authors remark that the background-reduction approach based on large quantities of lead is unfeasible for telescopes near the summit zone of Mt. Etna, where harsh conditions limit the access to heavy equipment. 
Another muography team~\cite{Catalano:2015zxd,DelSanto:2017ytw} has been active on Mt. Etna, with a very different detector technology based on Cherenkov light detection (see Section~\ref{sec:other-detectors}). The main advantage of a Cherenkov detector is the negligible background contamination (Section~\ref{sec:bkg}), that compensates the drawback of less statistics due to the larger intrinsic momentum threshold ($p > 5$~GeV, to be compared with the maximum at around 4~GeV of the atmospheric muon spectrum). 
This method is being developed with the existing ASTRI-Horn telescope, a prototype built in the context of the CTA (Cherenkov Telescope Array) project for astrophysics~\cite{CTA}. ASTRI-Horn is located at Serra La Nave, 5~km from the South-East Crater, but the authors advocate the construction of dedicated movable telescopes~\cite{DelSanto:2017ytw}, in order to reach different observation points.

\vspace{0.25cm}

Besides volcanology, the interest of muography has already been explored in several other Geoscience applications.  
These include the monitoring of groundwater and saturation levels for bedrocks in landslide areas~\cite{Azuma2014}, fault lines~\cite{Tanaka2011,Lesparre2016}, hydrogeological rock density perturbations~\cite{Hivert2017}, river banks damaged by animal activity \cite{Pazzi_EGU2019}, ice-filled cleft systems in steep bedrock permafrost~\cite{Ihl:2010uv}, 
and carbon capture storage sites~\cite{Klinger:2015gva,Gluyas2018}, as well as the exploration of natural caves~\cite{Caffau1997,Olah2012} and searches for minerals~\cite{Malmqvist1979,Schouten2018,Lingacom2018}. 
Being impossible to make justice of so many applications, we decided to elaborate only on a very recent development related to glaciology, probably less known because of its novelty~\footnote{Precursors of this new research direction were the proposal for alpine permafrost studies suggested in Ref.~\cite{Ihl:2010uv}, and in some sense also the pioneering paper by George~\cite{George1955} already cited in the Introduction.}: the study of the bedrock profiles underneath alpine glaciers~\cite{Nishiyama2017,Nishiyama2019}, using nuclear emulsion detectors installed in three observation points underneath the targets in a railway tunnel, with the methodology described in Ref.~\cite{Ariga2018}. 
The AM measurement yields, for each detector, the average density $\langle \rho \rangle$ along a line of sight; therefore, the position of the ice-rock transition surface is derived from the formula $\langle \rho \rangle = \rho_{rock}\cdot x + \rho_{ice}\cdot (1-x)$ (following Ref.~\cite{Huss2013}), where $x$ is defined as the fraction of rock between the observation point and the surface of the glacier along the line of sight. In order to minimize the model assumptions, the average bedrock density $\rho_{rock}$ is extracted {\it in situ} from angular regions not covered in ice, and its measurement is validated by comparing it with a set of rock samples collected from near the detectors along the railway tunnels and from the surface. 
Reference~\cite{Nishiyama2017} provided the first application of the method by studying the Aletsch glacier, in the Central Swiss Alps, measuring the shape of the ice-bedrock interface up to a depth of 50~m below the ice surface. They found a parallel orientation of the bedrock with respect to the glacier’s flow direction, which implies that the ice has passively slid on the bedrock without sculpting it. 
The same team then studied the Eiger glacier~\cite{Nishiyama2019}, 10~km away from the previous target, from different observation points within the same railway tunnel, finding a breach ($600 \times 300$~m) within the accumulation area where strong lateral glacial erosion has cut nearly vertically into the underlying bedrock. This suggests that the Eiger glacier has profoundly sculpted its bedrock in its accumulation area. 
Remarkably, the muography measurements for these targets are more precise and less model-dependent than any of the data available from traditional methods.

\subsection{Archaeology and Civil Engineering}
\label{sec:archaeology}

\begin{figure}[!ht]
\centering
\includegraphics[scale=.25]{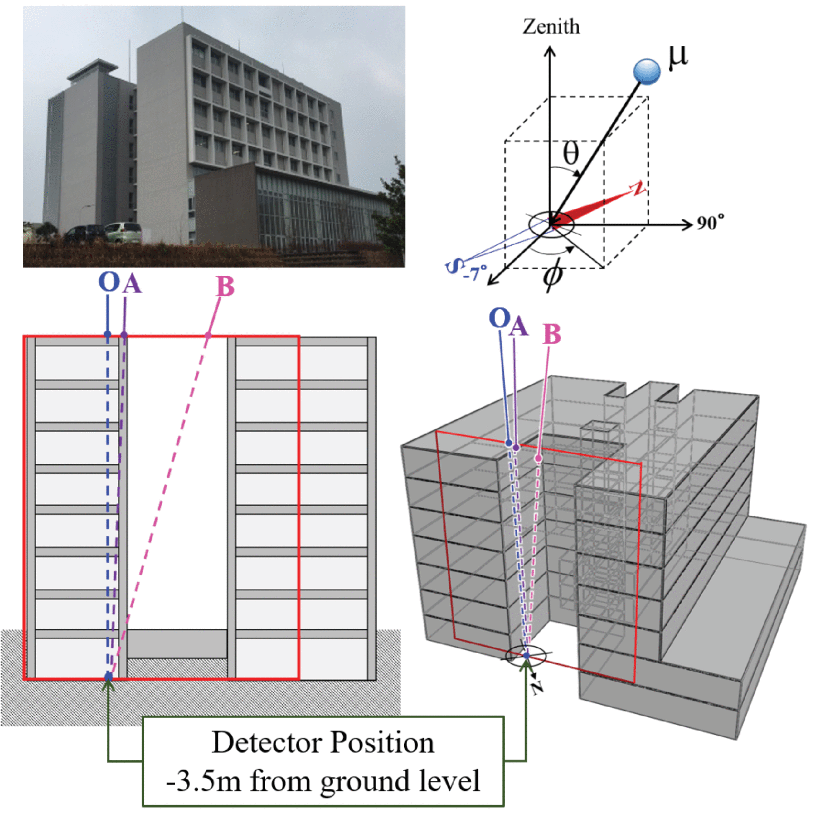}
\includegraphics[scale=.25]{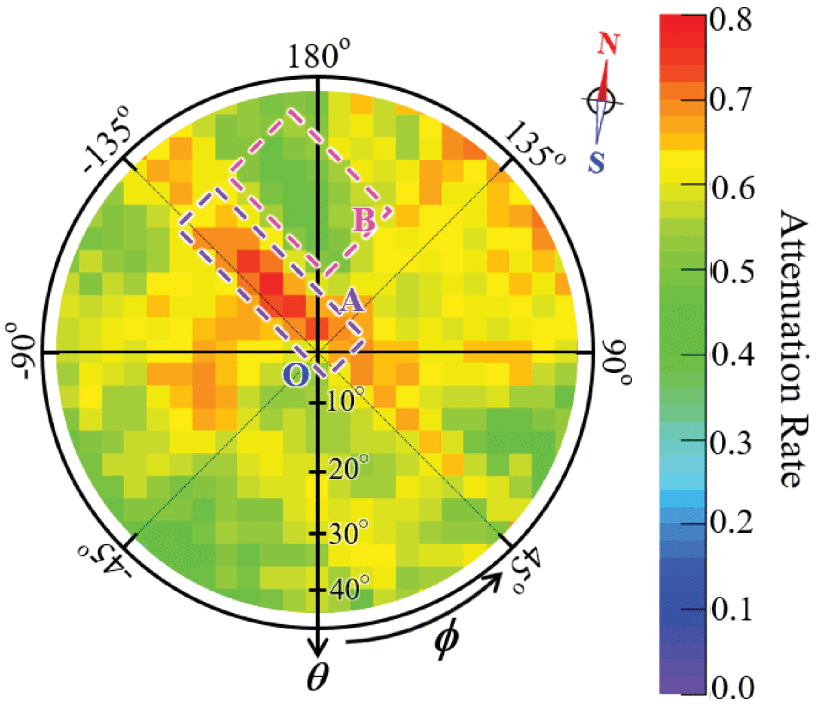}
\caption{Example of muographic imaging of a building. (Left) Building picture; sketch of its inner structure indicating the position of the detector (O) and the lines of sight of maximum (A) and minimum (B) integrated density; and definition of the $\theta, \phi$ coordinates. (Right) Muon attenuation map as measured from the detector in O. Reproduced from~\cite{Kyushu2018}.}
\label{fig:building}
\end{figure}

Applications of muography in archaeology and in civil engineering share many commonalities. 
In both cases the targets are man-made, and very often (but not always) the research questions that muography is called to address are related to absence or presence of voids. 
The typical size of the targets is such that AM is usually the method of choice~\cite{Gomez:2016kye,Marteau:2016jcn}; Fig.~\ref{fig:building}, from a recent study~\cite{Kyushu2018} with a portable muon telescope, illustrates how the muon flux attenuation can be mapped to the internal geometry of a complex building. 
However, some civil engineering applications in this area demand the ability to discriminate materials by $Z$, thus requiring the SM method; examples include the survey of the content of a blast furnace~\cite{Nagamine2005-furnaces,Vanini2018}, the study of reinforcement elements in the dome of Florence Cathedral Santa Maria del Fiore~\cite{Guardincerri:2016mrk}, and the measurement of the amount of wear suffered by a steel-made pipe~\cite{MuonSystems2018}. 


The very first muography application in archaeology has been Alvarez et al.'s imaging of Chephren's pyramid in Egypt~\cite{Alvarez1970}. 
Their study was inspired by the observation that the Second Pyramid of Chephren appears to have a much simpler internal structure than the Great Pyramid built by Cheops (or Khufu), Chepren's father; in general, the complexity of the internal architecture of the pyramids had an increasing trend during the Fourth Dynasty until the sudden appearance of simpler designs starting with Chephren.
The question that Alvarez et al. sought to answer was: are there unknown upper chambers of significant size in Chephren's pyramid above the Belzoni Chamber? Their data, compared to a Monte Carlo simulation of the expected muon flux, conclusively excluded that hypothesis. 

Decades later, the next pyramid to be surveyed with muography was the Pyramid of the Sun at Teotihuacan, Mexico, built by the Aztecs about 1800 years ago~\cite{Menchaca-Rocha:2014yxa,Melesio2014}. 
This pyramid is the third largest in the world, with a height of 75~m and a square base of $225\times 225~{\rm m^2}$. One of the motivations for this study was the search for inaccessible chambers that might hold the tomb of a Teotihuacan ruler. The detector was located in a deep underground chamber underneath the pyramid, accessible only through a tunnel 
so narrow that the muon telescope, $1.5~\rm m^3$ in volume and composed of six layers of multi-wire chambers (Section~\ref{sec:gas}), had to be dismantled and then reassembled inside. 
Data taking started in the early 2000s, and the preliminary results were released after more than a decade~\cite{Menchaca-Rocha:2014yxa}, reporting a very wide low-density volume in the Southern side, which has been interpreted by some researchers as an indication that the structure of the pyramid might have been weakened on that side and could be in danger of collapse~\cite{Melesio2014}. 
First steps towards muographic surveys have been made around 2010 also for the Mayan site of La Milpa in Belize~\cite{MayaMuonBook,MayaMuonWeb}, where the target is a tree-covered mound about 20~m high that is believed to hide a pyramid within; this is in fact, a typical case where standard remote-sensing technologies such as ground-penetrating radar cannot be used~\cite{Melesio2014}, as they require flat terrains free from rocks and roots to operate. 

\begin{figure}[!ht]
\centering
\includegraphics[scale=.21]{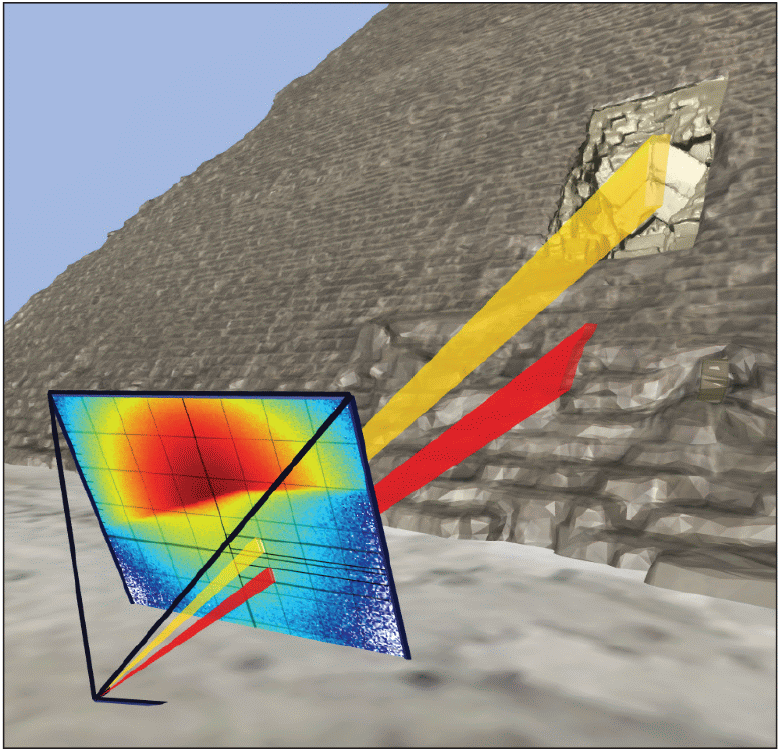}
\includegraphics[scale=.30]{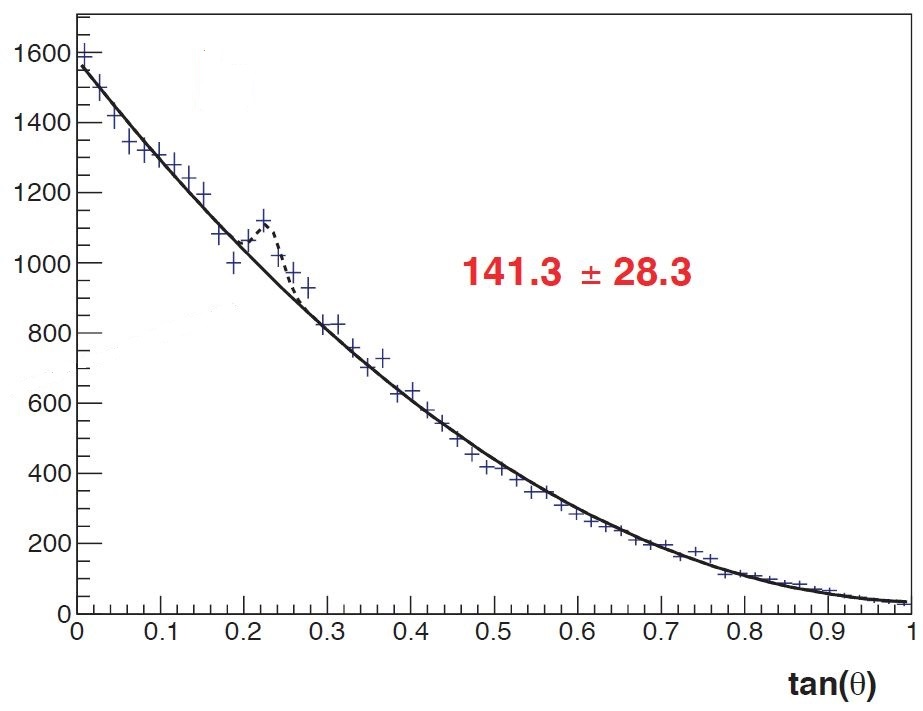}
\caption{(Left) Illustration of the correspondence between the image observed in an external detector and the face of the Great Pyramid; the yellow and red angular areas correspond to the unknown void and the Grand Gallery, respectively. (Right) Event counts (black points with error bars) as a function of the horizontal angle in the angular area indicated in yellow on the left. The solid curve is obtained from a model of homogeneous material. The peak corresponds to an unexpected excess of muons in the data. Reproduced from~\cite{Morishima:2017ghw}.}
\label{fig:scanpyramid}
\end{figure}

Muography came back to Egypt in 2015, in the framework of the ScanPyramids project~\cite{ScanPyramids} which combines several non-invasive techniques to survey Old Kingdom pyramids in search for unknown internal voids and structures. 
Three muography teams, each using one of the three main detector technologies that will be discussed in Sec.~\ref{sec:detectors}, have participated with simultaneous data taking. 
The highlight of the project has been the recent discovery~\cite{Morishima:2017ghw}, through muography alone, of an unexpected large void inside the aforementioned Great Pyramid of Giza, 
the oldest and largest in the Giza complex (139~m high and 230~m wide). 
Its three known chambers (known as the subterranean chamber, the Queen's chamber, and the King's chamber) are connected by several corridors, the largest being the Grand Gallery. 
The new void discovered by muography has a length of at least 30~m and a cross section similar to the Grand Gallery. 
The data were accumulated for several months with nuclear emulsions and scintillator-based telescopes installed in the Queen's chamber and two gaseous-detector (Micromegas, see Sec.~\ref{sec:gas}) telescopes located outside of the pyramid. This complementarity was useful for the 3D localisation of the void once the data from the three experiments were compared~\footnote{Anyway, for safety reasons, gaseous detectors would not have been allowed inside the tunnels by the Egyptian authorities. This is one of the frequent logistic limitations of this technique, as discussed in Sec.~\ref{sec:gas}. It may be remarked that this had not been an obstacle, instead, for the similar detector assembled underneath the Pyramid of the Sun~\cite{Menchaca-Rocha:2014yxa}.}; the emulsion detectors were positioned in two different locations, 10~m apart, allowing a stereoscopic image reconstruction with this method alone. 
All three teams reported an excess in muon flux originating from the same position in space, with statistical significance in excess of 5 standard deviations away from the null hypothesis (no void), as shown in Fig.~\ref{fig:scanpyramid} for one of the telescopes located externally. 
The expected excess of muons in the angular area corresponding to the Grand Gallery was used to validate the finding. 

An unknown cavity of potential archaeological interest has also been discovered within Mt. Echia, in Naples (Italy)~\cite{Saracino2017-Echia,Cimmino2019}. 
Mt. Echia is the site of the earliest settlement of the city of Naples in the 8th century BC. It is a headland with a maximum altitude of about 60~m above sea level and mainly consists of yellow tuff, a soft volcanic rock. In the course of history a very complex system of underground tunnels and cavities has been excavated and used for a variety of purposes, including the so-called Bourbon Tunnel that was excavated around the middle of the 19th century. 
First indications for the unknown cavity were reported in Ref.~\cite{Saracino2017-Echia}, based on the data from a 26 days pilot run with the MU-RAY telescope~\cite{MURAY2013}. The telescope, installed in the Bourbon Tunnel with a rock overburden of about 40 metres, had been oriented vertically and with a shorter distance between its planes with respect to its usage in volcanology. 
This result was confirmed, and the cavity more precisely characterized in 3D, by a second data-taking campaign whose results are reported in Ref.~\cite{Cimmino2019}. In the second campaign, the MU-RAY detector took data from a different observation point, and the MIMA portable telescope~\cite{Baccani:2018nrn} from a third location. MIMA can be tilted with respect to the vertical direction, and was oriented such to point towards the presumed location of the hidden cavity. 




\subsection{Nuclear safety and security}
\label{sec:nuclear}

Applications in this area typically involve SM, as they usually require to distinguish heavy elements from a background of lighter ones, and this was one of the prime motivations for the seminal paper~\cite{LosAlamos2003} that launched this method. In fact, the ability to distinguish between nuclear fuel (including spent fuel) and other metals is crucial for various applications of this kind that are of extreme societal relevance. 

Some applications require a very fast object detection (ideally less than a minute timescale), in order to have a minimal impact on queuing schedules. 
Examples include cargo inspections for homeland security, in particular the prevention of smuggling of nuclear material~\cite{LosAlamos2003, Riggi:2017izf,Lingacom2018}, as well as the search for radioactive material in scrap metal for the metal recycling industry, which is confronted with hundreds of contamination incidents per year, resulting in environmental issues as well as economic loss (a very expensive clean-up of the foundry itself is needed after each incident)~\cite{Vanini2018,Checchia2018}. 
Standard methods in these areas make use of so called ``radiation portals''; but these can fail if the radiation source is well shielded (in the first example, intentionally by the smuggler; in the latter example, by the scrap metal itself, or by a heavy-metal casing); therefore, ``muon portals'' are being proposed as a second line of defense, for further analysis of targets that are close to the alarm threshold~\cite{Checchia2018}: they have slower response than the radiation ones, due to the modest cosmic muon rate, but profit from the unique penetration power of the muon.
A few prototypes of such portals have been already built, typically using drift tubes (Section~\ref{sec:gas}) for muon tracking, although a recent proposal~\cite{Glasser:2018rhl} makes use of silicon microstrip detectors originally produced for the CMS experiment at the LHC. 
A muon portal is already in use since 2012 at the container port in Freeport (Bahamas) and is reported to have analyzed several thousand vehicles since its installation, most of them shipping container trucks~\cite{DecisionSciences}.

Muography is also being explored for safeguards applications related to the nuclear power industry. This includes the inspection of dry storage casks for spent nuclear fuel~\cite{Poulson:2016fre,Vanini2018} to verify if the cask has been tampered with, for example by determining if a bundle has been removed and either left empty or replaced with another dense material. 
This is another case where muography has an edge over measurements of $\gamma$ rays or neutrons from the fuel itself, as the cask walls are obviously designed to provide a very effective barrier against those kinds of radiation~\cite{Poulson2018}. 
Another application with many similarities in the area of nuclear safety is the imaging of the contents of nuclear waste containers and the quality assurance for nuclear waste treatment processes, as studied for example in Refs.~\cite{Ambrosino:2014aaa,Clarkson2015,Mahon2018}. 
In this kind of applications, the rapidity of response is not as crucial as in cargo inspections, so longer exposure times are acceptable. 
The size of the target is larger, meaning that larger detectors are needed, and that AM can be considered alongside with SM, as investigated in Refs.~\cite{Ambrosino:2014aaa,Vanini2018,Poulson2018}.

Finally, nuclear reactors themselves can be imaged with cosmic muons. 
Several studies~\cite{Borozdin2012,Miyadera2013,Kume:2016exx,Fujii:2019kwi} have been motivated by the nuclear crisis at Fukushima Daiichi (Japan) caused by a 9.0-magnitude earthquake followed by a tsunami in 2011. 
To this day, a radioactively contaminated area of 20~km radius around the nuclear plant may only be entered under government supervision. 
A question of huge relevance for the clean-up and decommissioning of the nuclear reactors is the location of the melted fuel, but direct access to the reactor buildings is hindered by radiation levels of order mSv/h, which motivates muography as a safe way to image the reactor cores from outside the buildings. 
At the same time, this also poses very specific challenges to the detectors, as the large flux of $\gamma$ rays from $^{134}$Cs and $^{137}$Cs induces a large number of individual hits that produce a significant combinatorial background, motivating studies of the optimal shielding thickness~\cite{Borozdin2012} and the development of novel time-coincidence logic circuits to minimize accidental coincidences of $\gamma$-induced hits~\cite{Kume:2016exx}. 
Uranium is at the same time one of the densest metals ($\rho = 19.1~{\rm g/cm^3}$) and the natural element with largest atomic number ($Z=92$); but fuel pellets contain uranium oxide mixed with other materials and sealed into zirconium alloy tubes, thus the average density of a fuel rod is only about $2.6~{\rm g/cm^3}$, which attenuates the muon flux only 2\% more than water. This poses a tough challenge when trying to distinguish the reactor core from the surrounding water through the building walls with AM. 
On the other hand, SM can achieve an image contrast of about 30\% with respect to water~\cite{Borozdin2012}, but it is not trivial to deploy large detector set-ups, that have a useful geometrical acceptance, in such a highly contaminated site. 
Results of muographic campaigns performed between 2015 and 2017 have been reported for some of the damaged reactors~\cite{WNN2015,WNN2017,Morishima-Fukushima}, using AM, showing that the fuel had melted and dropped from its original position within the core. 




%% file: detectors.tex
\section{Detectors for muography}
\label{sec:detectors}

Given the variety of applications, detectors for muography usually have to satisfy many requirements that are not always essential or even of interest in mainstream particle physics.
Typically, a muography particle detector should be rugged and capable of being operated remotely with minimal intervention. Power consumption can be a very important issue depending on the deployment site. As a bonus, the low event rate implies that the data rate and speed of the data acquisition (DAQ) and front-end (FE) electronics do not constitute a critical issue. In general, different muographic applications require different detector geometries and different detection technologies.
Thus, a large variety of muon detectors have been proposed and built during the last few decades, with only few common features between them. One of these is the inability of measuring the muon momentum on a per-particle basis, although this would be very desirable and possibly a major breakthrough for muography. 

\begin{figure}[t!]
\centering
\includegraphics[width=0.42\textwidth]{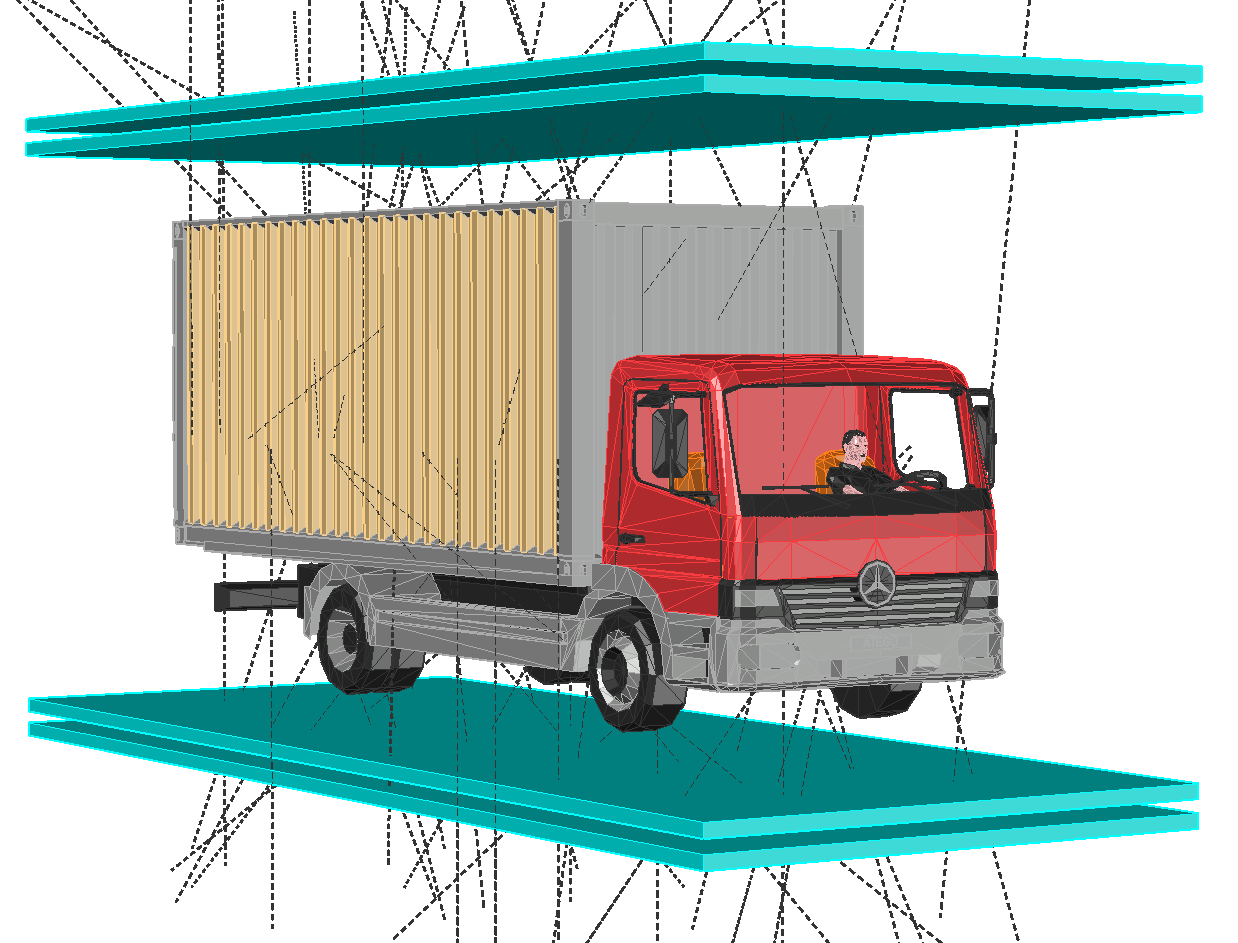}
\includegraphics[width=0.4\textwidth]{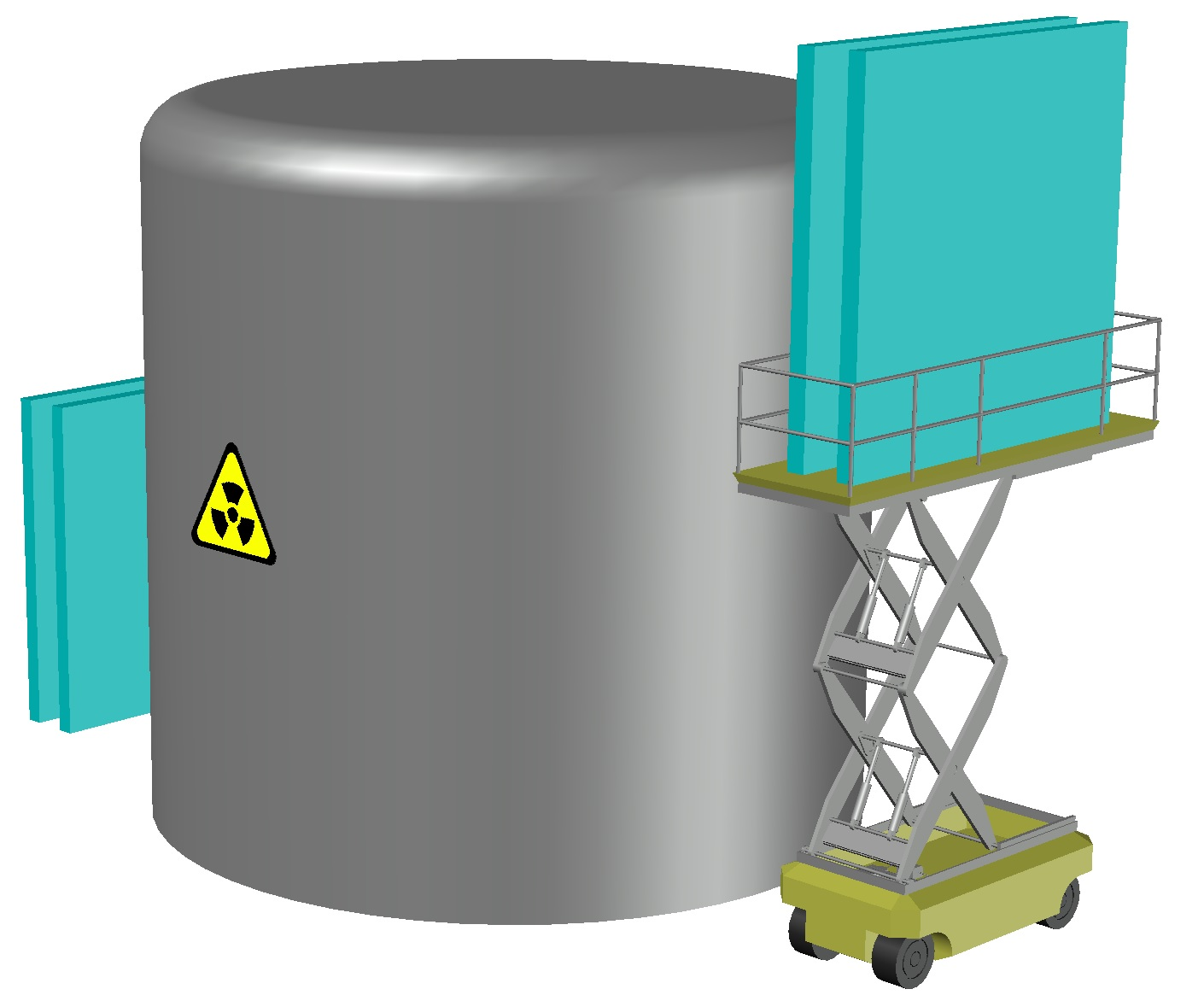}
\includegraphics[width=0.45\textwidth]{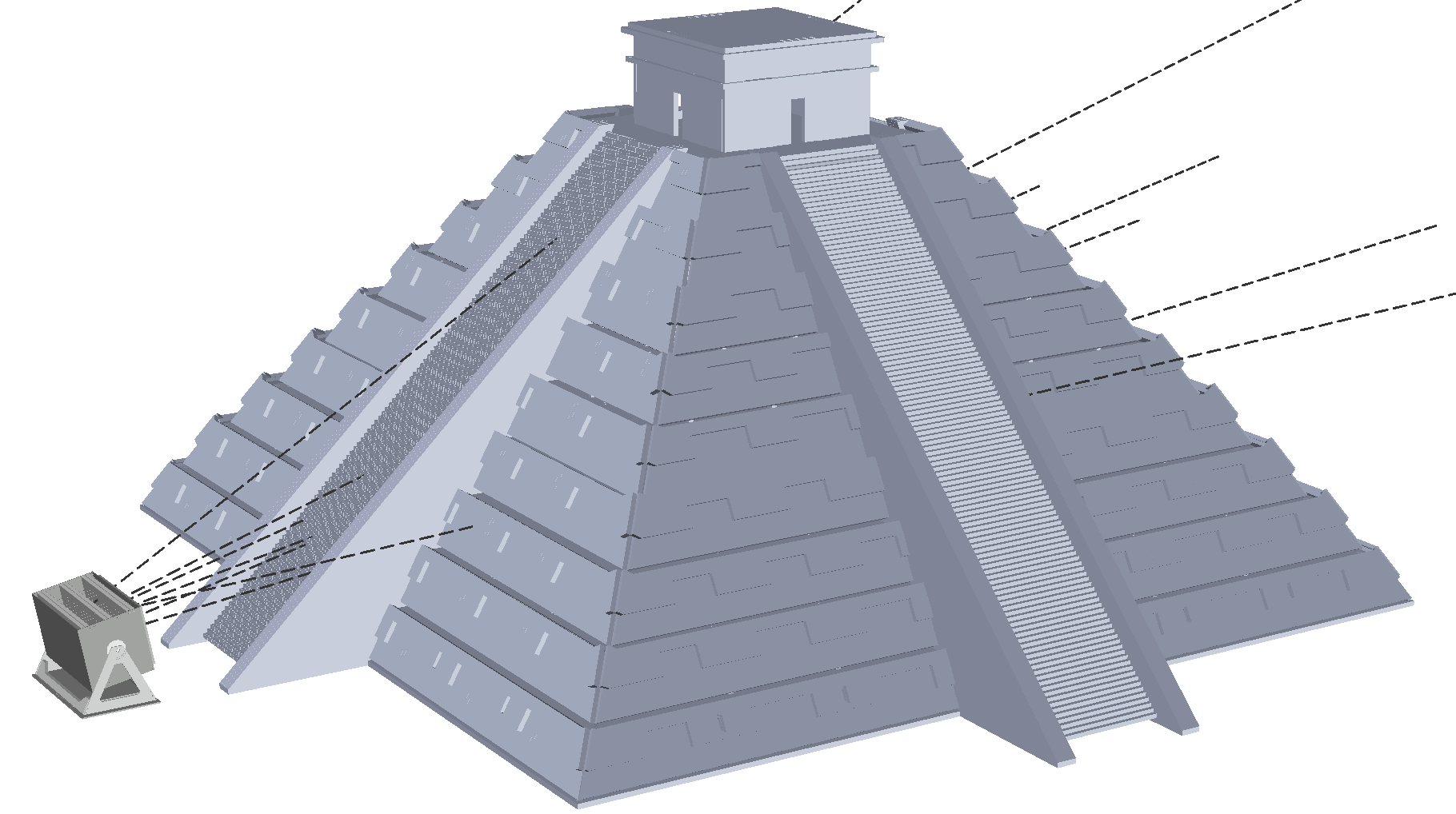}
\includegraphics[width=0.45\textwidth]{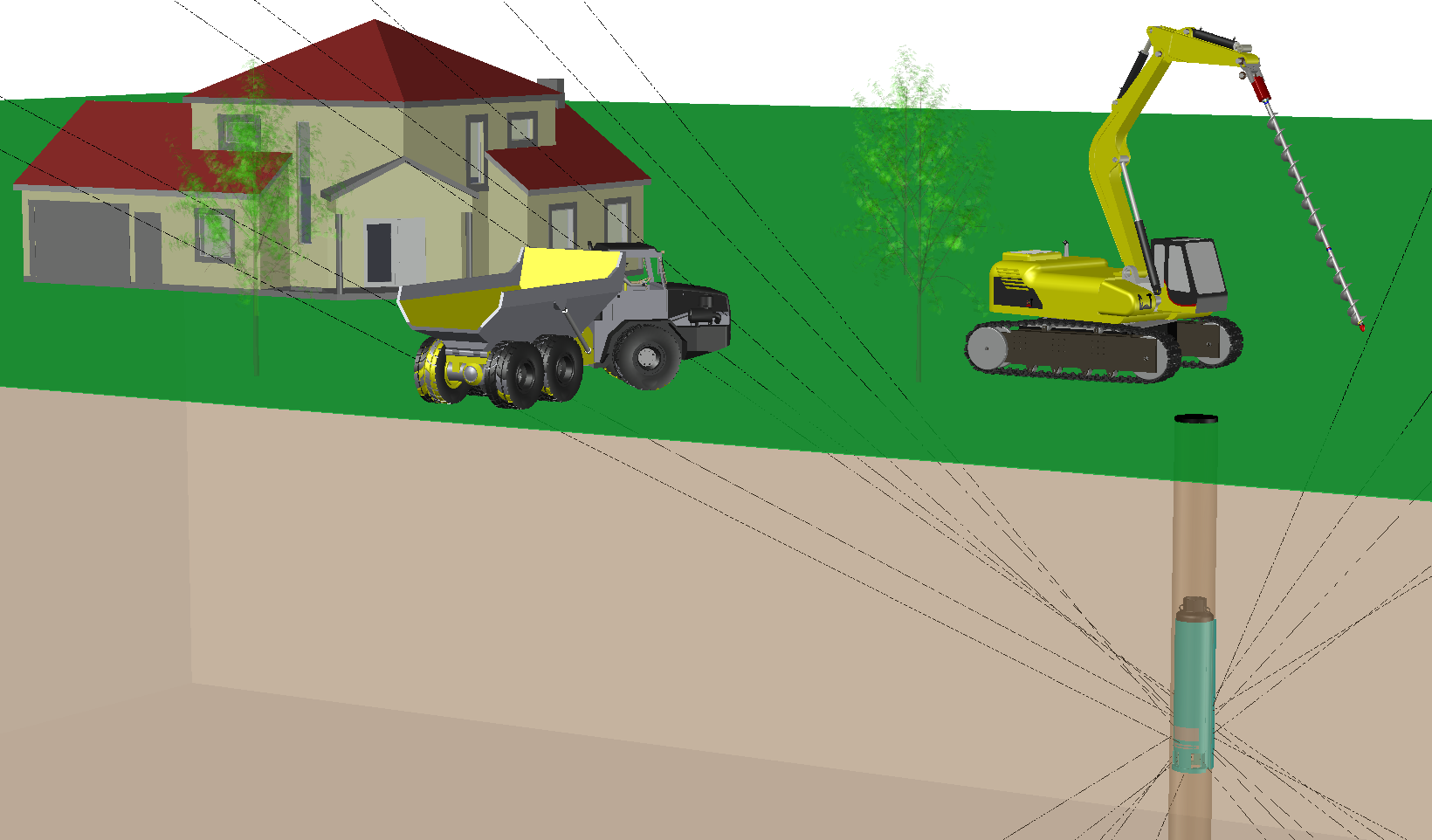}

\caption{Detector geometry depends on the application. A few examples are shown in the figure (clockwise from top left): cargo inspection, nuclear storage inspection, borehole application for underground imaging, and scan of a large open air-structure (e.g. a pyramid).}
\label{fig:common_detector_setup}
\end{figure}

If the detector system is intended for SM, the muon trajectory must be reconstructed with high accuracy (usually 1~mrad or better resolution) before and after the passage through the target being investigated. A typical detector geometry is sketched in Fig.~\ref{fig:common_detector_setup} (top left), with one of the two tracking systems above (upstream) and the other underneath (downstream) the target. This arrangement exploits the more abundant muon flux from the zenith, thus shortening the data acquisition time. This is the case for example of muon portals designed for homeland security, see Sec.~\ref{sec:nuclear}, where data must be acquired in the shortest time possible. 
However, the optimal geometry for specific use cases can vary, as shown for example in Ref.~\cite{Poulson:2016fre} where the upstream and downstream tracking systems are positioned on the sides of a large cylindrical cask of spent nuclear fuel (Fig.~\ref{fig:common_detector_setup}, top right). Due to the limited geometrical factor of such an ensemble, the detectors tend to be relatively large ($2-10~\rm m^2$) in order to maximise muon acceptance.

Where AM is the chosen technique, e.g. for imaging of large man-made structures ($\approx 10-100$~m), or very large targets such as volcanoes, a relatively small ($1-2\,{\rm m^2}$) detector can be placed laterally to the target, as sketched in Fig.~\ref{fig:common_detector_setup} (bottom left).  
The typical detector configuration is in the form of a \virgol{muon telescope} composed of position-sensitive layers, installed at a certain distance from the target. Since the multiple scattering angle is not measured, the angular resolution can be limited to 10~mrad or more. 
For a given X-Y spatial resolution of a single detection plane, the angular resolution of a telescope mostly depends on the distance between the first and last plane. A greater distance improves the angular resolution but decrease the acceptance of the telescope. 

When a muography is required of an underground target, this telescope geometry can only be exploited in the presence of tunnels of sufficient size below the target. When a suitable tunnel is not in place, given the high cost of drilling or excavating underground, absorption muography detectors have been proposed and built, that can  be inserted in boreholes (Fig.~\ref{fig:common_detector_setup}, bottom right). 
The first proposal of a borehole detector for muography appeared in Ref.~\cite{Malmqvist1979}, where simulations in a mine installation showed promising results that compared favorably to the telescope option.

The next sub-sections will focus on different muography detectors, classified by detection mechanism.

\subsection{Scintillation detectors}
\label{sec:scintillators}

Plastic scintillators constitute an ideal choice in many cases where spatial, and consequently angular, resolution is not crucial (as typical for AM).  Robust muon trackers, suitable for applications in harsh environments, based on fast-response plastic scintillator materials can be designed with a reasonable price/performance ratio ($\approx$ 50 K\euro$/{\rm m^2}$).

These detectors have become even more attractive since the advent of the Silicon Photo Multiplier (SiPM)~\cite{Hamawebinar}, replacing the traditional photo-multipliers (PM) at a much lower cost and power budget. 

\begin{figure}[t!]
    \centering
    \includegraphics[width=0.75\textwidth]{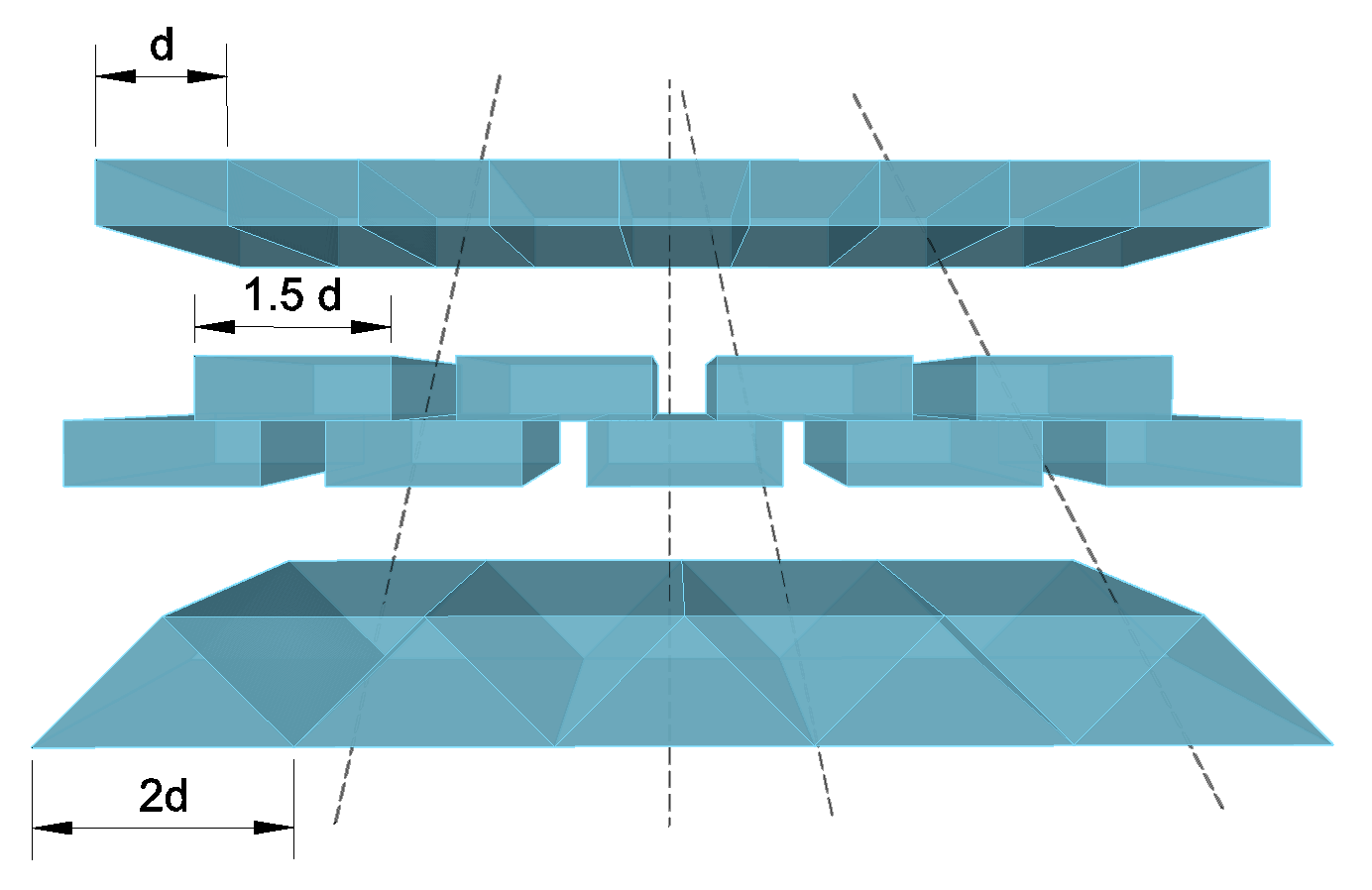}
    \caption{Three possible scintillator profiles for a position sensitive layer. (Top) Square or rectangular bars of the appropriate width are just placed side by side and readout digitally. (Middle) A variant of the previous design (tested in MIMA~\cite{Baccani:2018nrn}) that, with the same number of readout channels, increases the resolution by a factor two at the cost of a thicker (and more expensive) detection layer and of a more complex coordinate reconstruction. (Bottom) Triangular-section bars (as in MURAVES~\cite{Saracino:2017mao}) allow the use of a CoG algorithm to improve spatial resolution (that depends on scintillator light yield and FE electronics noise) with the same number of readout channels.}
    \label{fig:bar_shape}
\end{figure}

Plastic scintillators are easily shaped in various geometries, from square to rectangular to triangular bars for instance, that can be used to obtain a position sensitive detection layer with a relatively low number of readout channels. In general, they allow a remarkable customization of the detector’s geometry, as illustrated in Fig.~\ref{fig:bar_shape}. 
A relatively simple detection layer configuration is achieved using arrays of long plastic scintillator bars of rectangular or square section placed side by side to cover the required surface area, as in Fig.~\ref{fig:bar_shape} (top). 
Each scintillator bar is optically shielded from the others, so that a particle crossing produces a light signal in at most two adjacent bars, allowing a coordinate reconstruction. 
Figure~\ref{fig:bar_shape} (middle) illustrates how with a little ingenuity a plane can achieve a better spatial resolution with the same number of readout channels: in this example, introduced as a test configuration by the MIMA collaboration~\cite{Baccani:2018nrn}, the partial overlap effectively increases the spatial resolution whilst keeping the same number of readout channels, but using a more complicated coordinate reconstruction algorithm. The variance of the residuals distribution can be estimated as $L/\sqrt{12}$ for the first design (where $L$ is the pitch), while the overlapping layout, by exploiting the signals on adjacent bars, has a residual distribution variance halved with respect to the previous one. With both layouts, the residual distributions are flat and therefore the uncertainties are not Gaussian. For inclined tracks, the situation is more complicated. Nonetheless, an accurate coordinate reconstruction can still be achieved by using the track inclination information from other detector planes, through suitable algorithms.

As a third example, Fig.\ref{fig:bar_shape} (bottom) shows a configuration of bars with a right-angled triangular section. This solution, first adopted at the Fermi National Accelerator Laboratory (USA), has been used for the MU-RAY~\cite{MURAY2013}, MURAVES~\cite{Saracino:2017mao} and MIMA~\cite{Baccani:2018nrn} muography detectors. In this case, incident particles always cross two adjacent bars and the reconstruction of the particle impact coordinate takes advantage of the different amplitudes of signals produced in the two bars, roughly proportional to the track length in each scintillator. A simple Centre of Gravity (CoG) algorithm readily achieves a significant increase in resolution, better than that obtained with the second configuration layout (using the same number of readout channels), and with a Gaussian residual distribution. Unfortunately, the layout also affects significantly the weight and therefore the portability of the detector: for a given surface and number of readout channels, the second and third layouts weigh respectively 1.5 and 2 times more than the first.

Two detection planes are required for a tracking module with X and Y information. Multiple modules, placed parallel to each other, define a full three dimensional particle tracker. 
If at least $\rm N \ge 2$ tracking planes can be used, the resulting configuration not only mitigates possible backgrounds from spurious coincidence (see Sec.~\ref{sec:bkg}) but also has the capability of monitoring each plane detection efficiency.

\begin{figure}[t!]
    \centering
    \vspace{3mm}
    
    \includegraphics[height=3.5cm]{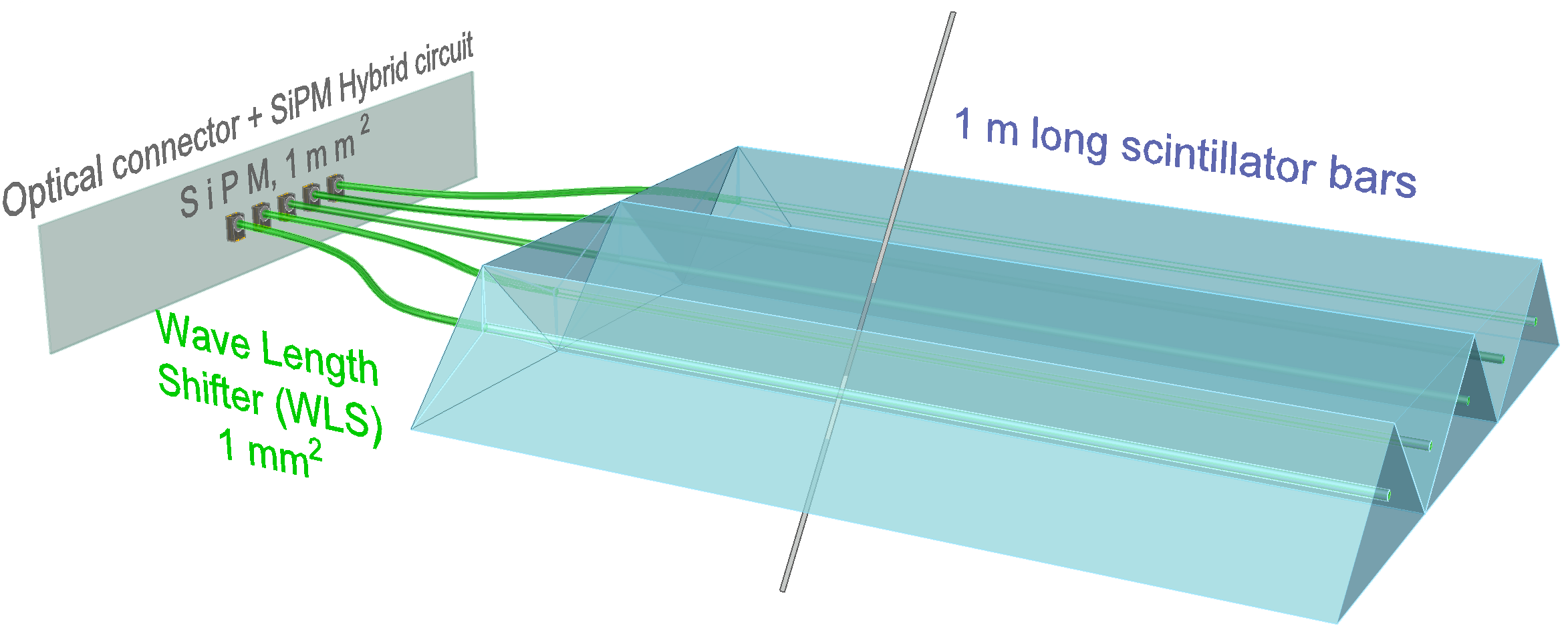}
    \hspace{0.5cm}
    \includegraphics[height=3.5cm]{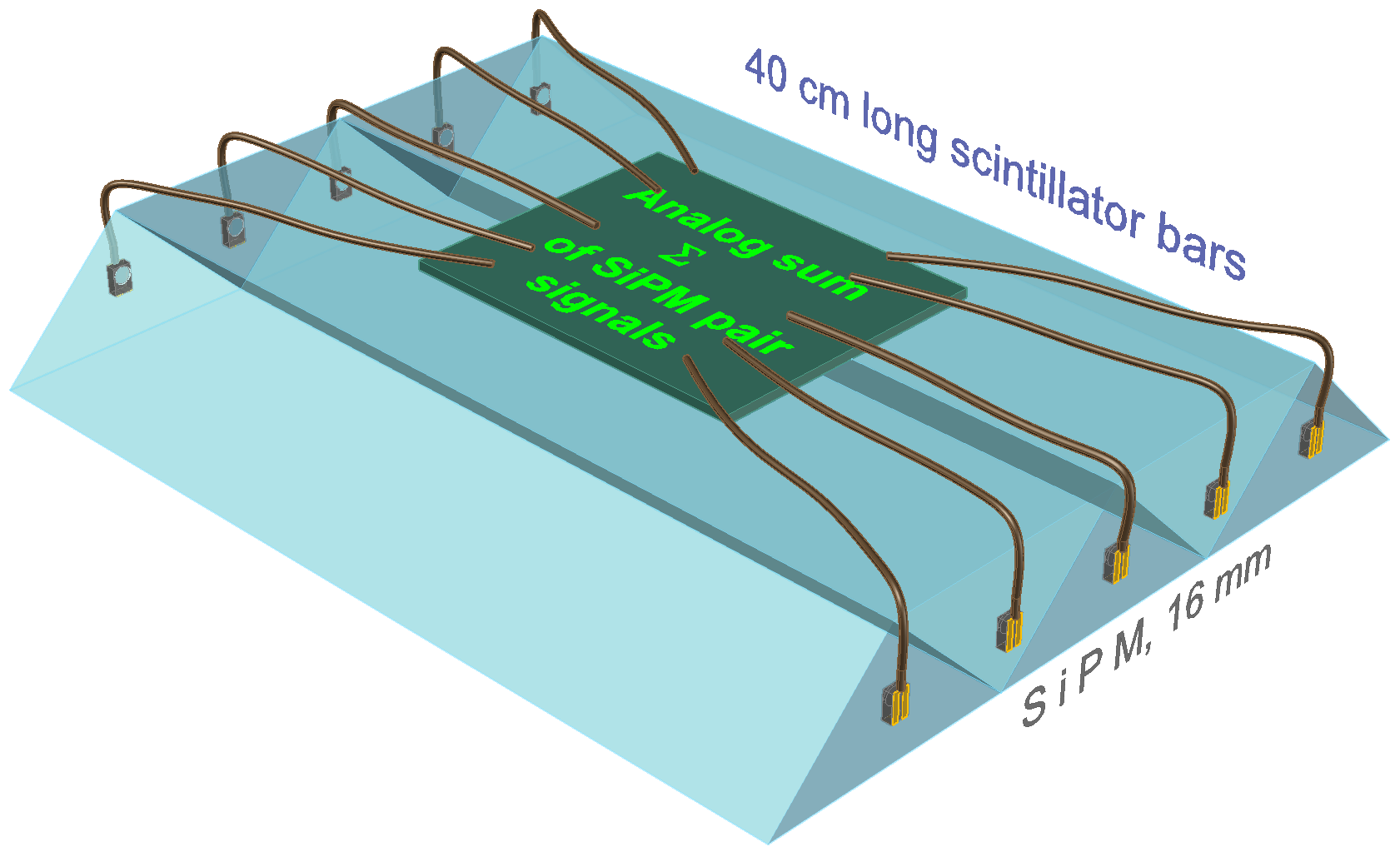}
    \caption{(Left) The WLS approach chosen by, for example, MU-RAY~\cite{MURAY2013}. (Right) Large SiPM approach used by MIMA~\cite{Baccani:2018nrn}.} 
    \label{fig:wls_mimasipm}
\end{figure}

Whatever the geometry, the scintillation photons emitted along the particle trajectory can be collected either through a light guide or through a wavelength shifter fibre (WLS) often  encapsulated in the scintillator bar itself~\cite{MURAY2013}, or with SiPMs (or PMs) directly coupled to the scintillator bars.
The WLS solution allows the use of very small ($\rm 1~mm^2$) SiPMs, thus lowering the costs of the FE electronics. On the other hand, large area ($9-16~\rm mm^2$) SiPMs are now readily available at a price only a factor $5-10$ times more than the smaller ones.  
Using these large area photo-sensors coupled directly to the scintillator (e.g. Ref.~\cite{Baccani:2018nrn} with triangular section bars) increases the number of collected photons, with a corresponding increase in the signal to noise ratio and in the achievable spatial resolution. In addition, there is no need to machine the scintillator bars for the WLS inclusion.
The design choices for two muography telescopes are illustrated in Fig.~\ref{fig:wls_mimasipm}.

A drawback of SiPMs is a marked temperature dependence of the breakdown voltage (typically $20-30 \rm mV/^\circ C$), which consequently affects gain, dark count rate and reverse current. Temperature conditions must be accurately and continuously monitored and the operating voltage changed accordingly. MURAVES~\cite{Raffaello2018} has implemented a thermo-electric (Peltier) cooling/heating module with an active controller capable of maintaining a constant temperature on the SiPMs, even in the presence of $\pm 10 ^\circ$C external variations. The power budget increase is of the order of 10 Watts for a $1~\rm m^2$ detection layer equipped with 64 SiPMs and two cooling/heating modules. A greater temperature compensation range (up to $\pm 20 ^\circ$C) can be obtained, simply increasing the power budget.

Plastic scintillators can also be shaped as thin fibres with square or circular section with typical transverse size of $1~\rm mm^2$~\cite{Borshchev2017}, greatly improving the spatial resolution at the cost of a much larger number of readout channels. Scintillating fibre planes are used in SM~\cite{Mahon2018}, coupled to multi-anode photo-multipliers (MAPMT) capable of reading out whole bundle of fibres with the necessary amplification.  Such a system has been developed and commercialised for nuclear waste management by Lynkeos Technology for use at the Sellafield storage site~\cite{Mahon2018}.

A particularly interesting development, that it is currently pursued by various groups, is the production of a fully functional independent muographic borehole detector system (e.g.~\cite{Bonneville2018,Gluyas2018}). 
New design configurations have been developed to maximize the angular acceptance of the detection systems~\cite{Battiston2006}, a must given the small radius of these detectors. In fact, borehole detectors have to meet stringent requirements in terms of compactness, ruggedness, impermeability, and performance in different environmental conditions. Ancillary equipment must also include some way of determining the detector orientation once inside the borehole. These detectors, although still in the development phase, could open up new application opportunities in mining and geotechnical surveys. 

\subsection{Nuclear emulsion detectors}
\label{sec:emulsions}
Nuclear emulsions are a special type of thick photographic plates with very uniform and fine sized (order $\mu$m) grain. Charged particles passing through a nuclear emulsion leave tracks with a spatial definition of the order of one micron, that can be seen at the microscope after developing the plates. Emulsions were among the earliest particle detectors and contributed to seminal results, like the discovery of the charged pion~\cite{LattesMuirheadOcchialiniPowell}. 
A recent large-scale application was the OPERA neutrino experiment, 
from where some recent spin-offs to muography~\cite{Morishima:2017ghw,Ariga2018,Tioukov2019} originated.

Nuclear emulsions maintain some peculiarities that make them the perfect solution for some types of applications. Their spatial resolution is of the order of microns, and multiple films of emulsions can be assembled to form thin tracking layers. These can achieve angular resolutions of the order of few mrad, have a limited cost and do not need any power supply~\cite{bozzaC-nucl_emul}. 
There are, however, important issues concerning their usage. First, the emulsion film starts recording particle tracks from inception. Thus an emulsion sandwich must be assembled right at the start of the observation campaign to \virgol{cancel} the information on the previously acquired tracks. In addition, they suffer from cold temperatures (i.e. below 10 $^\circ$C) and humidity.  
Another and very significant drawback is the equipment needed to analyse the plates. The OPERA experiment during its lifetime had invested considerable resources in the development of automated motorized optical systems that scan the plates and use pattern recognition to reconstruct track candidates in a reasonable amount of time (typically hours per cm$^2$). Only a few laboratories in the world are equipped with these microscopes thus limiting access to the technology. 

Nonetheless, emulsion films have been used successfully in a variety of muography experiments (see Sec.~\ref{sec:applications}) and have demonstrated an excellent performance in environments ranging from alpine tunnels~\cite{Nishiyama2017,Nishiyama2019} to hot sand deserts~\cite{Morishima:2017ghw} and  Mediterranean volcanoes~\cite{Tioukov2019}, during measurement periods spanning several months.

\subsection{Gaseous detectors}
\label{sec:gas}

A gaseous detector is often the ideal choice for applications where angular resolution is one of the main design parameters (which is always the case in SM). The muon crosses the gas volume leaving an ionised trail in its wake. The electrons (and ions) are collected by applying an electric field. Typical configurations use a cylindrical geometry, where the anode is a thin wire (of the order of 100 microns) that collects the electrons generated in the ionisation process. If the field is high enough ($50-100$~kV/cm or more at atmospheric pressure), a Townsend avalanche multiplication will occur in proximity of the wire. 
Gains of 10$^4$ or more are easily achievable while maintaining full proportionality between collected charge and initial ionisation. Since the signal can be quite large, the requirements on the FE electronics in terms of gain and noise are quite relaxed, greatly simplifying the design and reducing the cost. 
In general a gaseous detector takes advantage of the low CR flux using fewer, relatively simple, electronic channels (thanks to the gas amplification) and this in turn translates into high resolution, large surface detectors, at a cost which can be lower than the scintillator detectors described before. 

\begin{figure}[t!]
    \centering
    \includegraphics[width=0.3\textwidth]{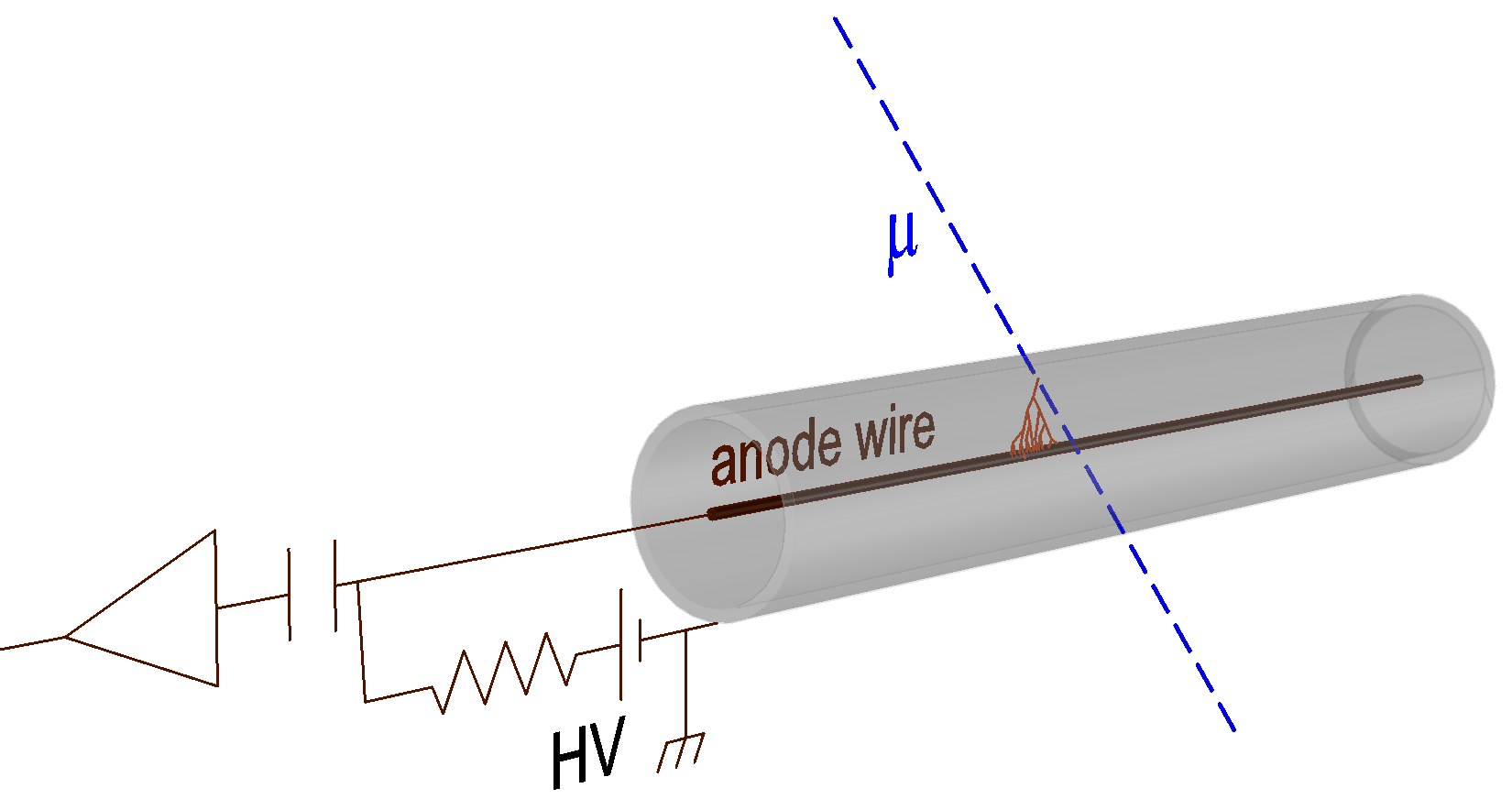}\hspace{5mm} \includegraphics[width=0.45\textwidth]{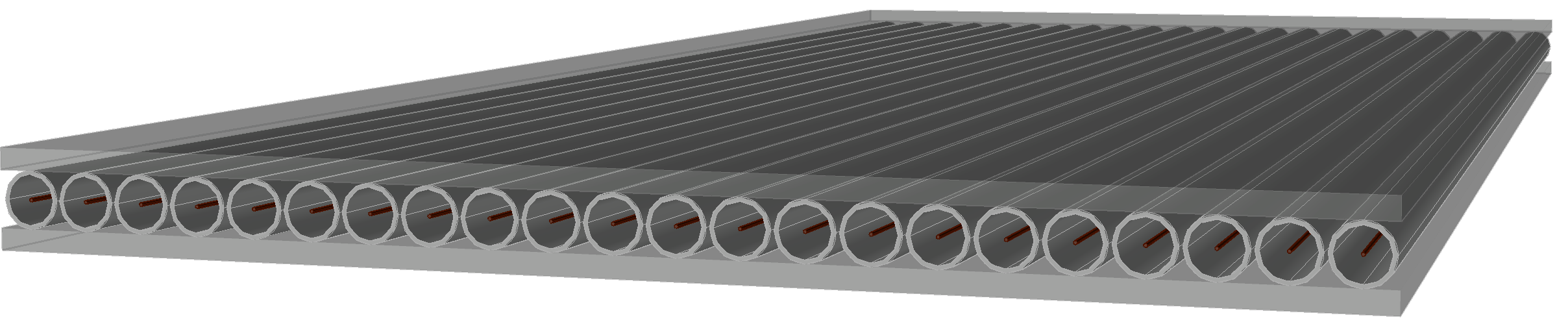}\\
    \centering
    \includegraphics[width=0.8\textwidth]{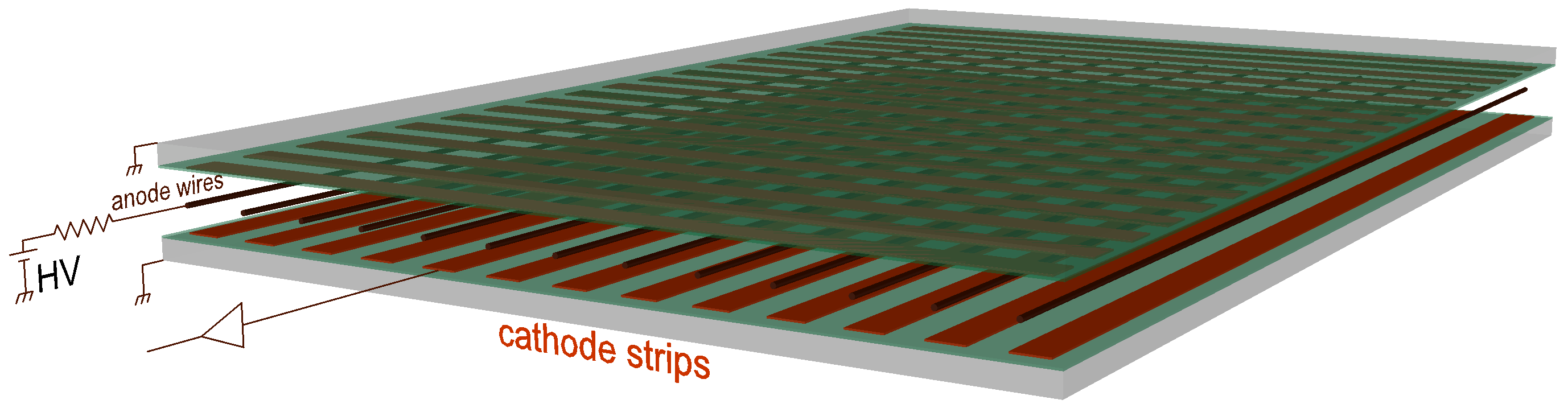}\\
    \includegraphics[width=0.85\textwidth]{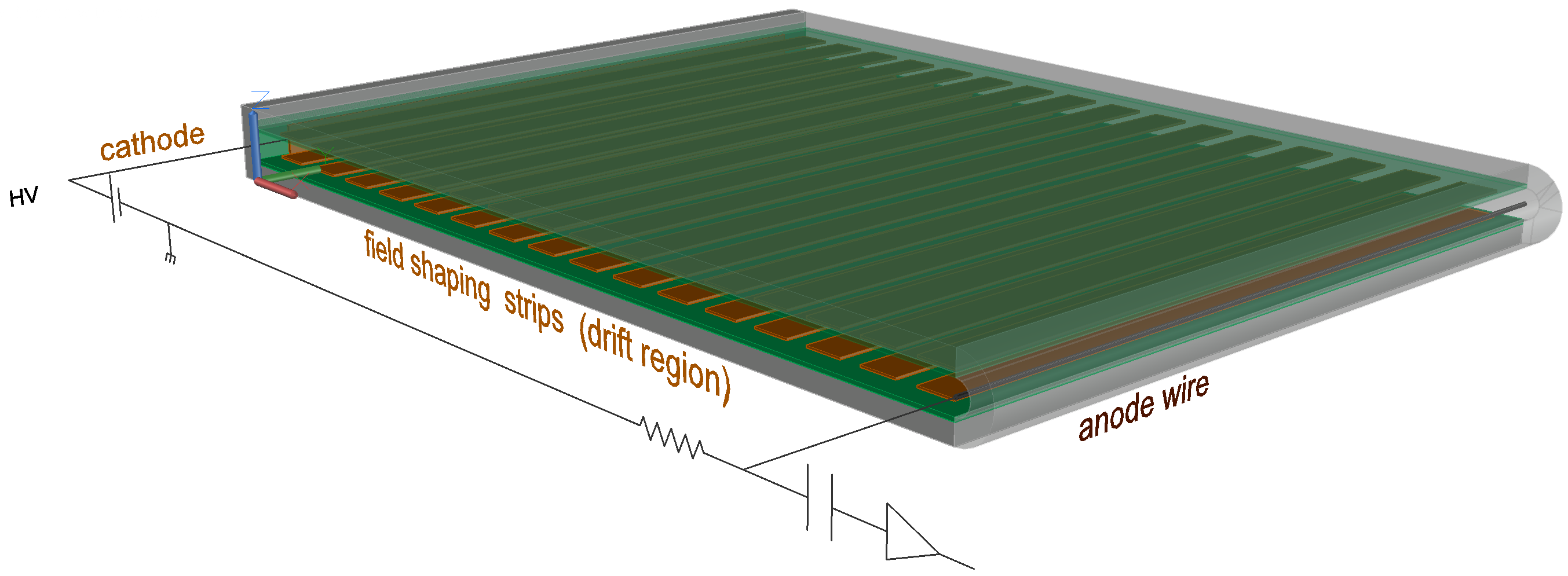}
    \caption{Principle of operation of a gaseous detector. (Top) A position sensitive layer is assembled from close walled cylindrical tubes. (Middle) An evolution of the previous design using a multi-wire approach that utilises segmented cathode strips to achieve a dual coordinate readout. (Bottom) Principle of operation of a drift chamber.}
    \label{fig:gas_assembly}
\end{figure}

A position sensitive detection layer using gaseous detectors can be relatively inexpensive to build (e.g. an assembly of aluminium tubes placed side by side). In the simple case of Fig.\ref{fig:gas_assembly} (top), the spatial resolution is given by the tube diameter divided by $\sqrt{12}$. The spatial resolution can be easily improved following a multi-wire chamber~\cite{Charpak} approach with segmented cathode strips for readout, see Fig.\ref{fig:gas_assembly} (middle). Simple CoG algorithms, using the charge induced on the strips, can easily achieve resolutions of a few 100 microns, at the price of a small increase in number of FE channels and detector complexity. 
Another approach, favoured by the low CR rate, is the use of drift techniques. A drift chamber, see Fig.\ref{fig:gas_assembly} (bottom), effectively measures the distance between the muon track and the anode wire, by measuring the time the electrons travel through the gas. The FE is more complex and Time to Digital Converters (TDC) are also required, but the number of channels can be drastically reduced, while obtaining resolutions of up to a 100 microns. This technique can be applied either with closed walled single tubes, or in a multi-wire chamber approach. Given the very high spatial resolution and consequent angular resolution, this type of detector has been used by various groups involved in homeland security projects~\cite{LosAlamos2014,Checchia2018}) or nuclear casks imaging~\cite{Poulson2018} using SM. 
Small-sized gaseous detectors can be also used in boreholes, as demonstrated by the studies reported in Refs.~\cite{OlahHamarMiyamotoTanaka2018,Lingacom2018}. 

Other types of gaseous detectors have also been used by groups involved in AM applications. Examples include Micromegas~\cite{Bouteille2016} in the ScanPyramids project~\cite{Morishima:2017ghw} and in geological prospections~\cite{LazaroRoche2016-micromegas}, and resistive glass gaseous pad detectors~\cite{Carloganu:2011zz} for the imaging of the Puy De D\^ome complex. In the Micromegas case, in order to contains the costs of the FE electronics, an innovative \virgol{genetic multiplexing} readout scheme~\cite{Procureur:2013yea,Bouteille:2016wdv} has been developed, thus preserving the inherent excellent space resolution of the detector while drastically lowering the number of FE channels. 
Micromegas are being also used in the design of a compact Time Projection Chambers (TPC) for geological prospections~\cite{Hivert2015}.
As a final example of gaseous detectors, the Resistive Plate Chamber (RPC) could in principle become a very economical large surface tracker. The detector works in avalanche mode, resulting in very large induced signals on the segmented readout pads. It is an economical detector, that requires a relatively simple FE electronics. A trial prototype~\cite{Baesso} for use in SM has achieved spatial resolutions of $300~\mu$m. 
Motivated by the low cost and relative simplicity of construction, a proposal has been made to also explore small-area versions of RPC for use in portable muon telescopes with gas-tight casings~\cite{Wuyckens2018}.

Notwithstanding the successes and the advantages of gaseous detectors, there are many issues concerning the operation of these detectors in muographic applications that are not under supervised laboratory conditions. The detectors usually need a continuous flux of gas, that translates in the need to supply and replenish gas bottles at the site where the measurements are made. Some detectors are relatively gas tight (no need to replenish for weeks or months) but in this case they are usually of small surface area. Some gas mixtures rely on the use of quenchers, explosive gases that pose significant safety issues in all underground applications. 
Moreover, drift velocity depends on the ratio of electric field to gas pressure which must be constantly monitored, together with temperature, and compensated for, also considering that the range of variation is much larger for muography \virgol{in the wild} with respect to the usual laboratory conditions.
In addition, these detectors are inherently more fragile (typical wire diameters are less than 100 microns) than scintillation ones and, in most cases, require a specialised laboratory with trained personnel to build and maintain the detectors. 
In particular large-area RPCs, while promising in many ways, have some specific drawbacks: they require a continuous flow of flammable gases, extremely high voltages ($\sim$10~kV), and can be permanently damaged by random sparking between the electrodes.

\subsection{Other detection mechanisms}
\label{sec:other-detectors}

Semiconductor position detectors (e.g. silicon micro-strips or pixels) have scarcely been considered for muography, in spite of their excellent spatial resolution (of the order of $10~\mu$m), because of their much higher cost with respect to the detector technologies outlined above. 
However, they may find their niche in specific applications where radiation hardness and compactness could be key factors, such as nuclear plants or highly radioactive waste management applications~\cite{Glasser:2018rhl}, or even space applications~\cite{Kedar2013}. 


The usage of the atmospheric Cherenkov imaging techniques for the imaging of volcanic conduits has been proposed in Refs.~\cite{Catalano:2015zxd,DelSanto:2017ytw}.
The physical principle at the basis of the method is the emission of electromagnetic radiation when a charged particle passes through a dielectric medium (such as air) at a speed greater than the phase velocity of light in that medium. 
The Cherenkov light produced along the muon path is imaged as a typical annular pattern (easily distinguishable from most fakes), and it contains enough information to reconstruct particle direction and energy. 
The natural momentum cut-off of the method is 5~GeV at sea level, a bit higher than the maximum of the momentum spectrum ($\approx 4$~GeV). 
A limitation is that data can not be collected in daylight. 
The construction of dedicated Cherenkov telescopes is very expensive, but where such an instrument is already present, e.g. for fundamental physics research~\cite{CTA}, it can be used parasitically for AM of nearby targets (see Mt. Etna's example in Sec.~\ref{sec:geophysics}).

\subsection{Summary of detectors for muography}
We conclude this section with a summary of various options in Table~\ref{tab:comp_det}.
\begin{table}[htbp]
\caption{\label{tab:comp_det} Summary comparison between different muography detector technologies.}
\smallskip
\begin{tabular}{|c|c|c|c|c|c|c|c|}
  \hline
  Type      & Surface & Resolution & Construction  & Readout & Cost & Suits & Applied Field \\
  \hline
  {\it Plastic Scintillators:}  &  &  &  &  &  &  &  \\
    Square Bars & 1-4 m$^2$ & $>$10 mrad & Simple & Simple & Low & AM  & A,G,V  \\
    Triangular Bars & 1-2 m$^2$ & $<$10 mrad & Simple & Simple & Medium & AM  & A,AT,G,GT,V  \\
    Scintillating Fibres & 1-2 m$^2$ & $\sim$0.1 mrad & Medium & Complex & High & SM & AT,GT,N  \\
 \hline
 {\it Gaseous Detectors:}  &  &  &  &  &  &  &  \\
    Proportional Tubes & 1-4 m$^2$ & $\sim$10 mrad & Simple & Simple & Low & AM  & A,G,V  \\
    Multi-wire Chambers & $>$4 m$^2$ & $<$1 mrad & Medium & Simple & Medium & SM  & AT,GT,N  \\
    Drift Chambers & $>$4 m$^2$ & $\sim$0.1 mrad & Complex & Complex & High & SM & AT,GT,H,N  \\
    Res. Plate Chambers & $>$10 m$^2$ & $\sim$0.1 mrad & Simple & Medium & Low & SM & AT,GT,H,N  \\
      \hline
  {\it Nuclear Emulsions}  & $<$1 m$^2$ & $<$10 mrad & Simple & Complex & Low$^*$ & AM & A,AT,G,GT,M,V  \\
  \hline
\end{tabular}
\\

\textbf{Legend:}\emph{ Low Cost $<$ 10K\euro/m$^2$, Medium Cost $<$ 50 K\euro/m$^2$, High Cost $>$ 50 K\euro/m$^2$;  AM = Absorption Muography, SM = Scattering Muography; A = Archaeology, AT = Architecture, G = Geology, GT = Geotechnical, H = Homeland Security, M = Mining, N = Nuclear Waste, V = Volcanology.}

$^*$ Excluding the automated scanning microscopes.
\end{table}

%% file: methods.tex
\section{Imaging}
\label{sec:methods}
Muography relies on measured muon fluxes as input data. As we already pointed out, the data collected by the apparatus provide accurate angular spectra but no direct information on muon momenta. Since cosmic rays have a markedly energy-dependent flux, this translates in the need to provide accurate \virgol{a priori} estimates of the expected fluxes in either clear sky cases or uniform material hypotheses.
This requires accurate Monte Carlo generators for cosmic ray spectra, that not only accurately reproduce the energy spectra down to a a fraction of a GeV, but also cover as much of the whole solid angle as possible. Also an accurate modelling of the outer profiles of the scanned object, whether it'd be a small cask or a volcano, must be provided as an initial input to the simulation software. 
Only then can a quantitative analysis be performed, which may span from a two dimensional (2D) density projection to a full three dimensional (3D) stratigraphy or tomography, depending on the data collected and the algorithms used.
The following sections describe what has been achieved in the field by the various groups involved. 

\subsection{Two dimensional imaging and density maps}
\label{sec:twoD}
Two dimensional imaging only concerns AM since, as we explain in the next section, SM naturally lends itself to three dimensional reconstruction. Track parameters are derived from the data collected, using the aforementioned trackers to reconstruct in space the trajectories of the detected muons. The track parameters are used to determine the angles defining the muon arrival direction, which are usually shown as a 2D muon angular distribution plot. Muon radiography requires that this distribution be measured not only downstream of the target object but also upstream (i.e. a \virgol{free sky measurement}), possibly pointing to exactly the same angular region. These measurements are usually performed with the same detector, to reduce the systematic effects due to the apparatus. The comparison of these two measurements, after correction for the duration of the data acquisitions and for the detector efficiencies, will result in a 2D angular map of the fraction $t$ of muons that are not absorbed by the scanned target. Usually this quantity, called muon transmission, is obtained from the ratio between the corrected angular distributions $N_{T}(\theta,\phi)$  and $N_{FS}(\theta,\phi)$, measured respectively downstream (T=target) and upstream (FS=free sky) of the target volume~\cite{Tanaka2009,Olah2012,Menchaca-Rocha:2014yxa,Morishima:2017ghw,Saracino2017-Echia}:
\begin{equation}
    \label{eq:muon_transmission}
    t (\theta,\phi)=\frac{N_{T}(\theta,\phi)}{N_{FS}(\theta,\phi)}
\end{equation}

To extract information on the density distribution inside the target volume, a comparison with a simulation is required. Simulations (see Sec.~\ref{sec:mc}) have to include a realistic muon flux describing the angular and energy spectra at ground level, the detailed external geometry of the volume under examination, as seen from the point of observation of the detector, and an estimated average value of the volume density.
Comparing the measured transmission with different simulations, each of them performed assuming different values of the average density, will result in a full 2D average density map distribution, as seen from the detector’s viewpoint~\cite{Temperino2019}. 

This analysis technique relies on a precise reconstruction of muon trajectories. The spatial resolution of the selected tracking technology is therefore a parameter of interest, because the corresponding angular resolution is one of the parameters that determine the accuracy of a density map. Common muon trackers developed for muon radiography have angular resolutions $\Delta \alpha $ going from few milliradians to few tens of milliradians. 
In those cases where the detector's size is very small compared to the object's size and distance, the spatial uncertainty $\Delta r$ due to the detector angular resolution, when describing a structure located at distance $L$ from the detector, can be simply calculated as $\Delta r=L\times\Delta \alpha$. Considering muography applications to volcanology (Sec.~\ref{sec:geophysics}), a typical situation is represented by $L\approx 1~{\rm km}$ and $\Delta\alpha \approx 30~{\rm mrad}$, implying $\Delta r \approx 30~{\rm m}$, better than many other geophysical methods commonly used in the study of the internal structure of volcanic buildings~\cite{Nishiyama2016}.\newline
Where this approximation is no longer valid (e.g. most archaeological applications), each point of the target is seen at different angles from different parts of the detector and an additional uncertainty in the spatial reconstruction of the target object, of the order of the detector size, has to be considered. 

An important source of uncertainty is related to multiple Coulomb scattering suffered by muons while traveling through the target before being recorded by the detector. The effect builds up as muons loose energy through matter. In fact, the energy spectrum of muons exiting from a typical target is peaked at low energy and, without a momentum cut-off, the largest fraction of muons detected by a normal muon tracker will peak at momenta around 1\,GeV/$c$. This has motivated various groups to use iron/lead filters in their muon telescopes (see Sec.~\ref{sec:momentum}).

\subsection{Three dimensional imaging: stereoscopic reconstruction, back-projection, and full 3D reconstruction}
\label{sec:threeD}

\begin{figure}[t!]
    \centering
    \includegraphics[width=0.8\textwidth]{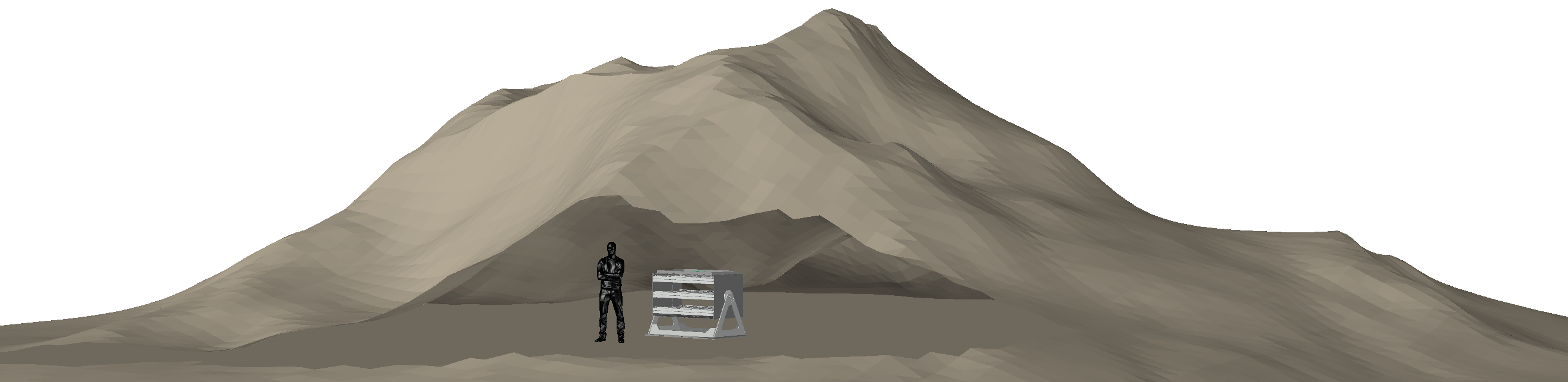}\vspace{3mm}
    
     \includegraphics[width=0.8\textwidth]{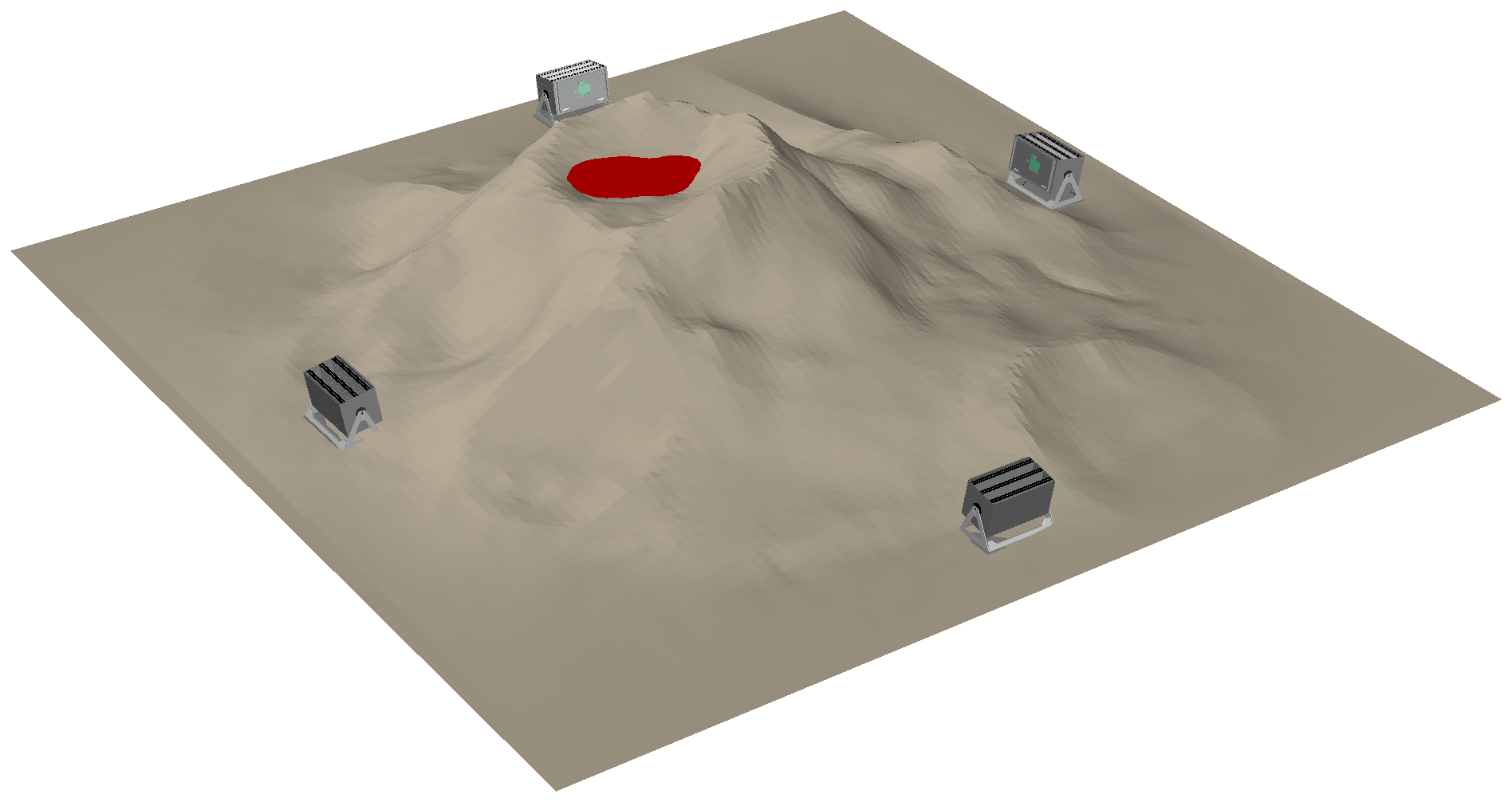}
   \includegraphics[width=0.7\textwidth]{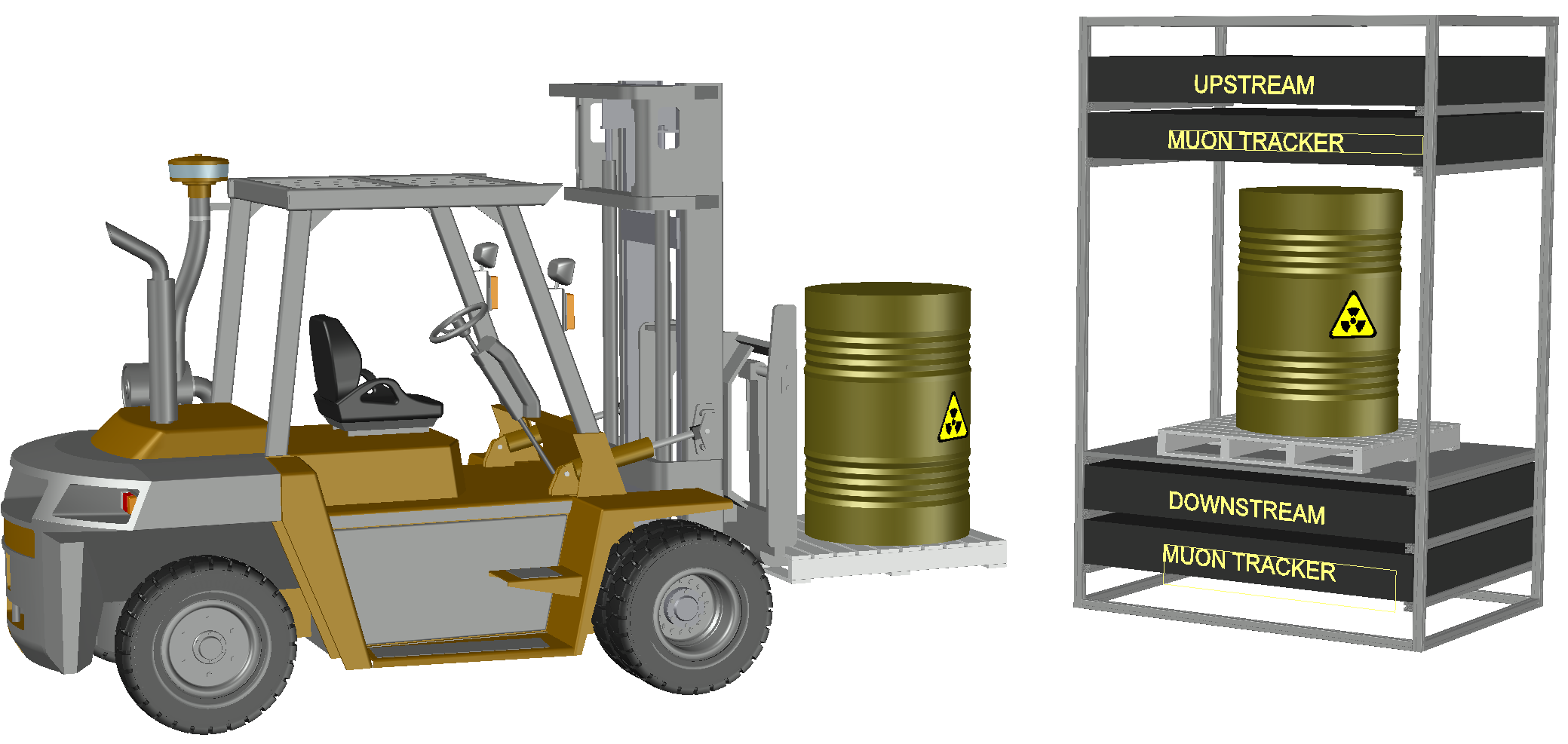}
 
    \caption{Examples of muographic measurements aimed at a 3D reconstruction. (Top) AM applied to an overburden inside a tunnel. (Middle) AM applied to a volcano conduit. (Bottom)  Typical SM application with the target placed between two detection layers.}
    \label{fig:3d_msm_and_am}
\end{figure}

In general, data from more than one viewpoint can be correlated to infer a 3D distribution of the scanned target, either observing the target from different viewpoints simultaneously, or moving a single tracker around the target. The latter approach minimizes cost and logistical complexity, but it assumes that no evolution on the timescale of the measurement can affect the target or the detector or the environmental conditions. 
In Fig.~\ref{fig:3d_msm_and_am} (top and middle) we show two typical configurations with AM telescopes, while for SM we point back to Fig.~\ref{fig:common_detector_setup} where a portal for homeland security (top left) and a scan of a nuclear cask are shown (top right); Fig.~\ref{fig:3d_msm_and_am} (bottom) shows an example where a small nuclear waste cask sits in between two tracking detectors.

Typically the object is sliced in a three dimensional voxel matrix, with assigned initial densities. An iterative minimisation procedure varies the voxel densities till a best fit to the data is obtained. This is a classic inversion problem, where the solution is not univocally determined, and where increasing the constraints on the algorithm, using data from many angles (i.e. with more measurements), can often lead to instabilities and failure to converge to a solution. It is outside the scope of this review to provide a treatise on this topic, suffices to say that the regularisation algorithms that have been developed can mitigate many of the issues and are based either on minimum squares formulations~\cite{Tarantola1982,Press1992} or Bayesian ones~\cite{Tarantola2005}. 

Both AM and SM techniques can use this approach, with spectacular results.
There are numerous examples of AM usage to obtain three-dimensional density images of volcanoes. One of the first examples~\cite{Tanaka2010} used data collected from two nearly orthogonal vantage points at Mount Asama in Japan. The technique was then integrated, see the end of this section, with gravimetry data to reconstruct the inside of the Showa-Shinzan lava dome~\cite{Nishiyama2014}. Other examples of 3D reconstruction of volcanoes with AM can be found in Ref.~\cite{Tanaka2015,Marteau:2015pxa,Lesparre2012,Carloganu2013}. Other applications of AM in 3D imaging have measured tunnel overburdens~\cite{Bryman2014, Tanaka2015}, in particular for mining exploration~\cite{Schouten2018}. Remarkably, the Los Alamos group~\cite{Guardincerri2017} obtained a detailed stratigraphy of the tunnel overburden, including the top soil layer, with a four point measurement, using only muographic data. 
In one particular case~\cite{Cimmino2019}, a search for a hidden cavity (described in Sec.~\ref{sec:geophysics}) was conducted using an innovative algorithm that did not rely on an inversion procedure or a full 3D reconstruction, using instead a clustering algorithm that combines the information from three viewpoints to identify a volume of very low density in space, with a high statistical significance. 

As mentioned before, imaging by a single AM set-up is naturally two-dimensional. 
However, recent work~\cite{Bonechi2015back_projections} has shown how to derive 3D information with AM from a single observation point, when the detector's size is not negligible compared to the distance and size of the target object. The back-projection algorithm developed by the authors, is based on the fact that different sides of the detector see the target from different angles, and so a stereoscopic view of the target is in fact present in the data. The difference with standard 2D AM is that the algorithm exploits not only the information on the direction of arrival of muons, that is used for the reconstruction of muon angular maps, but also the measurement of the impact points on the detector~\cite{Bonechi2018UK}. 
This method is best suited for locating volumes characterized by densities that are very different from the surrounding material, like empty cavities or metal deposits inside a rocky layer. Although the resolution along the rock depth is modest (about one meter in the conditions of Ref.~\cite{Temperino2019}), the back-projection algorithm has the merit of requiring a single measurement site, unlike conventional 3D methods that can become impractical or impossible to execute at certain locations (e.g., when the detector can only be located within a narrow tunnel~\cite{Saracino2017-Echia}.)

As stated before, three dimensional imaging is a natural output of SM observations. 
First implementations~\cite{Morris2003,LosAlamos2003} used a Point of Closest Approach (POCA) algorithm to determine the voxels densities. In this method only one scattering centre within the scanned target is allowed. Albeit somewhat brutal, this approximation yields very good results~\cite{Chatzidas2016} with 3D image resolutions of a few mm$^3$. 
More sophisticated algorithms make use of the so called maximum likelihood expectation maximization (MLEM)~\cite{Clarkson2015,Checchia2016, Checchia2018}, where multiple scattering centres are allowed, and even combine scattering information with absorption data~\cite{Vanini2018}, using simulations to estimate the expected muon flux. While some targets are small enough to be scanned in one go by the apparatus, measurements on nuclear casks~\cite{Poulson:2016fre}, given the size of the targets, require multiple measurements that are anyway beneficial in resolving potential ambiguities and can improve image resolution.


\vspace{0.25cm}
We now consider the use of collateral measurements to aid the 3D imaging process. 
This mostly applies to AM applications in geosciences, archaeology and civil engineering (see Sections~\ref{sec:geophysics} and~\ref{sec:archaeology}), where several geophysical tools are available for surveys, with complementary merits to each other and to muography. 
For volcano imaging, typical sources of additional data include gravimetry, seismic noise, seismic waves, and geoelectric data; 
all of these methods need to rely on strong model assumptions to solve the degeneracies of their strongly non-linear inversion problems. Muographic data though, being directional, naturally break those degeneracies. 
Ground penetrating radars (GPR) and magnetometers are popular tools in near-surface surveys, including archaeology. 
In some SM applications for relatively small targets, X or $\gamma$ ray picture profiles can provide useful starting information. 

Gravimetry is special in this context, as its combination with AM comes natural given that both methods are directly sensitive to the rock density distribution. A gravimeter is an instrument designed to measure the local value of the acceleration of gravity, $g$, which is affected by the nearby mass density distribution. 
Given the spherical symmetry of the gravitational field, a single gravimeter has no directionality, and is only sensitive to distance (the effect of an object of density $\rho$ and volume $V$ at a distance $r$ from the gravimeter is given by $\Delta g = G\rho V/r^2$, where $G$ is Newton's constant), therefore gravimetry campaigns are based on arrays of measurements. 
The inverse dependence on $r$ also causes the method to be increasingly less sensitive to increasingly distant volumes of interest; while this is to some extent true also for muography (large target depths reduce the available statistics), its key advantage is that its instrumentation does not need to be installed on the surface of the target itself. 
Successful examples of combinations of gravimetric and muographic data are given in Refs.~\cite{Nishiyama2014,Jourde2016,Rosas-Carbajal:2017sdu,Lelievre2019-joint}. 
Usually gravimetric data are used as initial seeding for the 3D voxel density matrix, that is then refined using the muographic data. 
Reference~\cite{Gonidec:2018kij} shows an example where the various data were analysed independently and only combined in the final stages of the analysis, eliminating possible biases but considering the correlations between data sets at the end of the analysis process.


Sometimes, different standard methods can produce wildly different results. For example gravimetry and seismic tomography have given discordant results about the deep structure of Mt.Vesuvius~\cite{Cella2007,Denatale2006}. In this case muographic data can help determining the correct solution to the inversion process. Similarly, traditional methods (such as petrology, ground deformation, seismicity, gas geochemistry, etc.) hint at the presence of a large dyke at the interior of Mt. Stromboli~\cite{Calvari2014}  but are unable to provide accurate estimates of its dimensions; therefore, a muographic image with a target resolution of 10 metres or less would have a very significant impact.

In general any 3D imaging, whether using SM or AM, benefits greatly from any initialisation data that constrains the voxels densities. 
Of course, adding more measurement methods translates in additional costs for the measurements and for the instrumentation needed. For example, the cost of a gravimetric measurement campaign can escalate in the tens of K\euro, without considering the price of the gravimeter itself. On the other hand, in some cases these complementary data are already available.

We conclude this section by remarking that muography has shown to be a very robust imaging technique that can survive on its own, especially if one can increase the number of observational points (i.e. telescope viewing angles). Moreover, in many cases, muography has a larger potential for breakthrough improvement (with a relatively small investment) with respect to traditional methods that are, at most, in the incremental improvement stage.

%% file: issues.tex
\section{General issues for muography}
\label{sec:issues}

This section introduces some issues that affect a broad range of applications of the method, and on which significant ingenuity is invested by the muography practitioners, that are rarely addressed in reviews of this topic.

\subsection{Background reduction and estimation}
\label{sec:bkg}

Two types of background are of concern for most muography use cases: ``fake'' muons and ``soft'' muons. 

\begin{figure}[!t]
\centering
\includegraphics[width=.45\textwidth]{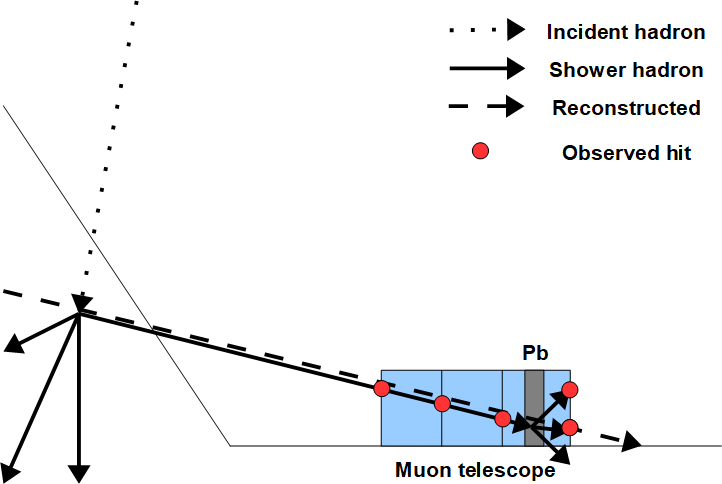}
\hspace{1cm}
\includegraphics[width=.45\textwidth]{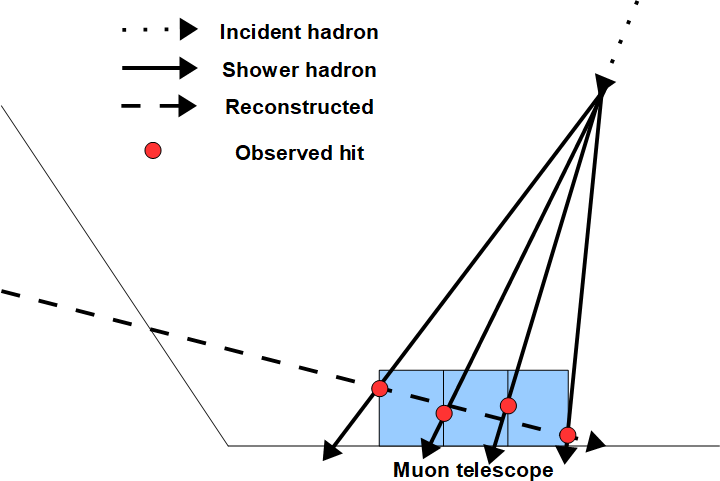}
\par \vspace{1cm}
\includegraphics[width=.45\textwidth]{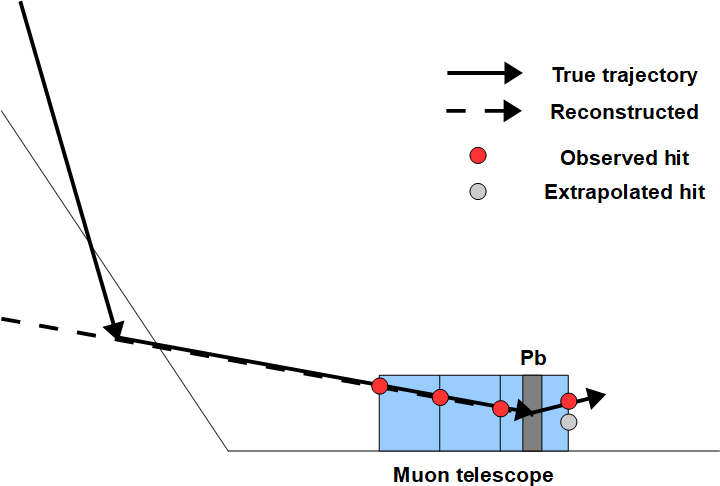}
\hspace{1cm}
\includegraphics[width=.45\textwidth]{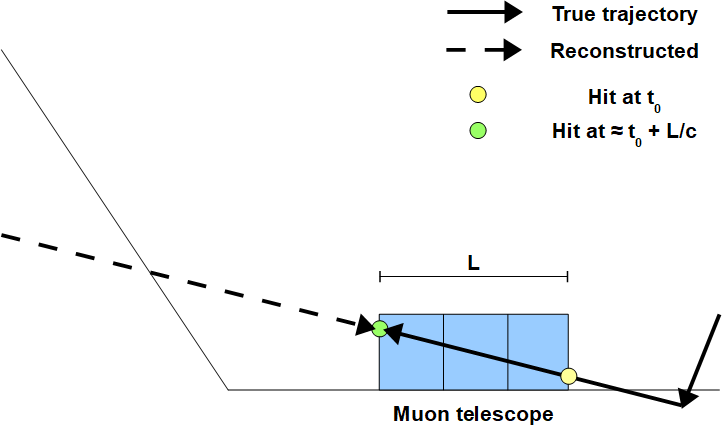}
\caption{(Top left) Fake muon, e.g. a charged hadron. (Top right) Combinatorial background. (Bottom left) Soft muon. (Bottom right) Backward muon. Background mitigation includes the use of lead to cause further inelastic interactions (top left) or further elastic scattering (bottom left), strict selection on the quality of the reconstructed track (top right), and usage of hit-level timing information (bottom right).}
\label{fig:bkg-muons}
\end{figure}

Fake muons are charged hadrons or $e^\pm$ (see for example the studies in Refs.~\cite{Nishiyama2016,Olah2017,Olah:2017zgo}, based on detailed Monte Carlo simulations and validated with data) that manage to reach the sea level in spite of their large interaction cross sections, typically because produced late in the cosmic-shower development, see for example Fig.~\ref{fig:bkg-muons} (top left). 
Their abundance, composition and spectrum at sea level depend on the same factors that affect the flux and momentum spectrum of real muons (see Sec.~\ref{sec:momentum}) and therefore their precise determination is location-dependent as well. For this reason, and also because they tend to concentrate at the lower side of the spectrum where Monte Carlo models are less reliable and reference data are scarce, it is particularly important to estimate them \emph{in situ}.  
The term fake muon is sometimes used in the literature to also indicate combinatorial background, i.e. tracks composed by random associations of hits that do not originate from a single particle, as in Fig.~\ref{fig:bkg-muons} (top right). 
This is easily reduced by a modest redundancy in the number of layers and an upper limit on the $\chi^2$ of the track fit. The cosmic flux at sea level is sufficiently low with respect to typical data acquisition times that random coincidences of aligned hits from distinct particles can usually be neglected; exceptionally, however, combinatorial background may be induced by several temporally correlated particles, for example those produced from the same cosmic shower~\cite{Saracino:2017mao}.

Soft muons are actual muons with relatively low momenta (as opposed to \virgol{ballistic} muons, defined as those whose speed can be approximated with the speed of light). They are a nuisance for muography because the probability of a large-angle scattering depends on the inverse of the momentum, as shown in Eq.~\ref{eq:scattering}. Thus soft muons, cause a blurring of the image~\cite{Gomez:2017omj} with consequent loss of details, see Fig.~\ref{fig:bkg-muons} (bottom left). 
This is a particular concern for very large-scale (i.e. \virgol{thick}) targets, such as mountains and volcanoes, as the observing telescope is necessarily distant and oriented quasi-horizontally. In this situation the muon flux becomes very low and the scattered soft muons can easily become dominant after large-angle scattering in the surrounding ground. 
A related concern, when a muography telescope has to be oriented horizontally or quasi-horizontally, is that muons can enter from its rear (either ballistic muons produced at very low elevation angles, or soft muons scattered from the ground behind the detector~\cite{Carbone2013}), as in Fig.~\ref{fig:bkg-muons} (bottom right). Like before, in the presence of very low fluxes, if no mitigating action is applied, these ``backward'' muons may even overwhelm the muons who carry information about the target. 

A very effective passive method for reducing the fake background, already exploited since the seminal study by Alvarez et al.~\cite{Alvarez1970}, is to include an absorber (e.g. an iron or lead slab) in the detector setup: hadrons and $e^{\pm}$ have large probability to undergo a destructive interaction, respectively nuclear or electromagnetic, and either no hit is found after the absorber, or more than one hit is found (signalling the presence of a nuclear or electromagnetic shower, respectively). 
This method also reduces the number of soft muons, in fact from equations~\ref{eq:dedx} and~\ref{eq:scattering} an effective momentum cut-off is introduced by the absorption in the passive material, while the remaining soft muons tend to get a large kink in their trajectory, allowing to identify them in the offline analysis~\cite{Saracino:2017mao}. 
The latter effect is better exploited by high-resolution detectors: in nuclear emulsions (Sec.~\ref{sec:emulsions}), with their excellent angular resolutions, thicknesses of a few millimeters of iron between the emulsion layers are sufficient to detect a kink, while scintillator-based telescopes need thick walls of lead in order to achieve the same rejection factor, with a significant adverse effect on the detector portability.
As an example of the latter case, the optimal trade-off between signal efficiency and background rejection for the scintillator-based MURAVES detector is achieved with a 60~cm thick lead wall between the last and next-to-last layers~\cite{Raffaello2018} in order to get a momentum cut-off of around 1~GeV (corresponding to a $\approx 20\%$ attenuation in the overall muon flux. 
This is not a problem \emph{per se}, as most of these are unwanted soft muons, but the detector mass increases from roughly 100 Kg to more than five tons.

Particle-identification techniques can help to reduce some types of fake muons. 
At low momentum, the specific energy loss of $e^\pm$ and protons is larger than for muons, and this can be exploited via the pulse height of the signal, if digitized with a sufficient resolution and properly equalized across the detector channels. The same principle is also employed in the $\Delta E - E$ method for particles that stop in the detector (Sec.~\ref{sec:momentum} and Ref.~\cite{SatoKinWatanabe2017}). 
This technique is not effective in rejecting pions, however, due to their similarity in mass to the muons leading to a similar $dE/dx$. 

Backward going muons can effectively be rejected by measuring the time of flight (TOF)~\cite{Marteau:2013qua} of the detected particles between the front and the rear layer of the telescope~\footnote{In some set-ups employed in the muography literature, dedicated scintillator layers with excellent time resolution are installed, for the purpose of high-resolution TOF measurement, before and after the position-sensitive detectors.}: a ballistic muon, having $v\approx c$, employs 3.3~ns to travel 1~m, meaning that a timing resolution of O(ns) is needed for quasi-horizontal telescope orientations, easily reachable with current detectors and FE electronics. 
If the telescope can be rotated, the TOF measurement can be precisely calibrated under the reasonable assumption that all detected muons enter a vertical telescope from the top. 


Finally, we remark that the Cherenkov detector principle described in Sec.~\ref{sec:other-detectors} has negligible background contaminations. 
Soft muons are obviously not an issue because of the high intrinsic momentum threshold, and fake muons are easily rejected from the shape of the Cherenkov emission pattern. Also the background due to back-scattered muons from the ground has been estimated to be negliglible ($\approx 3\times 10^{-3}$ fake events per night)~\cite{Catalano:2015zxd}.  

\subsection{Muon momentum}
\label{sec:momentum}

As both processes of interest for muography (absorption and scattering) happen in any case, and both depend on momentum (Eqs.~\ref{eq:dedx} and~\ref{eq:scattering}, respectively), they constitute a nuisance to each other: in absorption-based muography the large-angle scattering of low-momentum muons causes a blurring of the images; while in scattering-based muography the absorption of muons while traversing the target causes not only a loss of statistics (usually not a big concern for small- and medium-size targets) but also a selective bias. In both cases, correcting for these nuisances introduces additional model dependencies. 

An accurate description of the momentum spectrum depends on the location~\cite{Grieder2001,Cecchini:1998kv}: altitude is obviously important as the shower development and the energy loss in the atmosphere depend on the amount of air above the observer; and the Earth's magnetic field causes dependence on latitude and longitude of the observer and direction of observation through the geomagnetic cutoff~\footnote{Charged particle trajectories are bent in presence of a magnetic field, therefore the intensity of the local geomagnetic field determines the minimum momentum necessary for a charged particle of cosmic origin to reach the ground.}; also local anomalies in the geomagnetic field can have a visible effect on the spectrum of charged cosmic particles~\cite{Hebbeker2002}. Reference data for atmospheric muon spectra have been tabulated at various locations~\cite{Grieder2001}, but the spectra also vary with time~\cite{Grieder2001,Cecchini:1998kv}:  the solar activity influences very significantly the flux and energy spectrum of primary cosmic rays in both periodic and aperiodic ways, and the pressure in the atmosphere also plays a role: short-term variations in the lower atmosphere due to the weather change the density of air traversed by muons before reaching the ground~\footnote{Weather can also affect the muon flux by electromagnetic effects during thunderstorms, as measured in Ref.~\cite{Hariharan:2019fai}, where the muon flux variation versus time provides direct evidence for the generation of potentials of order $10^6$ V in thunderclouds.}, and seasonal pressure variations in the upper atmosphere due to the temperature affect, for the same reason, the mean free path of pions and kaons and therefore their probability to decay into muons before undergoing nuclear interactions~\cite{Adamson:2009zf}. 
Temperature effects are expected to distort the momentum spectrum of the muons~\cite{Hebbeker2002} because the momentum of their mother particles (pions and kaons) correlates with their lifetime via relativistic time dilation~\cite{Grashorn:2009ey}.

Some muography teams have produced their own reference data, using dedicated detector setups. For example, several Italian muography studies make use of the tabulated data from the ADAMO magnetic spectrometer (developed originally as a test prototype of the magnetic spectrometer of the PAMELA satellite experiment)~\cite{Bonechi:2003bi}. Compact and portable, and equipped with a 0.4~T permanent magnet, ADAMO measured the all-particle differential flux as a function of momentum (from 100~MeV to 100~GeV) and zenith angle (from $0^\circ$ to $80^\circ$)~\cite{Bonechi2005}.

In alternative to a magnetic spectrometer, the absorption of muons in volumes of known material can be used to select a momentum range. 
This approach has been followed for example by a Japanese team~\cite{SatoKinWatanabe2017} to measure the muon flux in momentum bins from 50 to 130~MeV, in an acceptance of $40^\circ$ around the zenith. 
They used a simple apparatus composed of a top detector used as a trigger, a bottom detector used as a veto, a thick middle detector where the low-momentum muons are stopped, and a slab of lead to act as a momentum selector. The bottom veto ensures that only muons that stop in the middle detector are analysed. Under these conditions, because all of their kinetic energy is deposited in the middle detector, the energy distribution of the stopped muons is directly measured. Background from $e^\pm$, particularly important at low momentum, is strongly reduced with the $\Delta E - E$ method~\footnote{This method, based on Eq.~\ref{eq:dedx}, exploits the fact that at low momentum the specific energy loss of $e^\pm$ is much larger than for muons.}, exploiting the energy measurements in the top ($\Delta E$) and middle ($E$) detectors. 
A similar idea is also at the core of the recent  NEWCUT facility~\cite{NEWCUT}, operated in Japan by an Hungarian-Japanese team for the purpose of measuring reference spectra over a broad momentum range and at various zenithal angles. 
The NEWCUT rotating telescope is composed of several detector layers alternated with lead slabs. The length of a muon track through the telescope, hence the amount of lead that it penetrates, is directly related to its initial momentum. For energetic muons that are able to punch through all the lead slabs, momentum is estimated from scattering via Eq.~\ref{eq:scattering}.

Good reference data can be used directly (in look-up tables or parametrized) or indirectly (by tuning Monte Carlo generators for cosmic ray showers, see Sec.~\ref{sec:mc}) to improve the precision of the measurements. However, ideally one would like to measure the momentum of individual muons \emph{in situ} in order to minimize model dependence. 
Doing that with a magnetic spectrometer would be impractical or too expensive for most muography applications, but a method to get indirect access to momentum is illustrated for example in Refs.~\cite{Checchia2018,Vanini2018} in the context of SM. 
The basic idea (similar to one of the soft-muon removal methods discussed in Sec.~\ref{sec:bkg}) is to reconstruct the muon trajectory, after having traversed the target, before and after some iron slab whose purpose is to cause further muon scattering. 
One can then exploit the known size and material of the slab, and the $\sigma(\Delta\theta)$ after the target as measured in the bottom part of the apparatus, to extract the average $1/p^2$ distribution by approximate inversion of Eq.~\ref{eq:scattering}:
\begin{equation}
    \langle 1/p^2\rangle \propto \frac{X_0}{x}\cdot\sigma(\Delta\theta)^2 \, .
\end{equation}
However, directly feeding $\sqrt{\langle 1/p^2\rangle}$ as a replacement for $1/p$ in Eq.~\ref{eq:scattering} is too crude, as $p^2$ can vary by a factor $10^4$; the solution found in Ref.~\cite{Vanini2018} is to extract $\langle 1/p^2\rangle$ in sub-sets of data defined by the $\chi^2$ of the track fit, exploiting the strong correlation between $p$ and $\chi^2$. Furthermore, as discussed in Ref.~\cite{Checchia2018}, it is important to take into account the bias on the $1/p^2$ distribution due to absorption, that removes from the sample of scattered muons the lowest-momentum ones. This effect depends on $x$, creating a deviation from linearity that needs to be estimated from Monte Carlo simulation.

We point out that in general, any kind of muon momentum determination that can be implemented in a muographic detector is not only useful for background reduction but would also greatly improve the signal significance. Even a rough, threshold based, momentum knowledge of the muon both before and after the target would be an enormous aid to any SM muography~\cite{Vanini2018} given the dependence on momentum of multiple scattering and the relevance of relatively low momentum muons to these type of measurements. We surmise that the same holds also for AM, but typical applications (e.g. volcano surveys) would require the capability of measuring muon momentum up to at least several tens of GeV, which would probably require a magnetic spectrometer in conjunction with a high resolution detector.

\subsection{Monte Carlo issues}
\label{sec:mc}

Comparing data with Monte Carlo simulations for the observables of interest is crucial for the imaging of a target. 

The simulation chain starts from the generation of muons with realistic angular and momentum distributions. Solutions range from {\it ab initio} simulations that simulate the entire cosmic shower in the atmosphere (including, therefore, the background particles), such as CORSIKA (COsmic Ray SImulations for KAscade)~\cite{CORSIKA}, to parametrizations of the muon distributions of interest based on existing data (e.g., the dedicated measurements mentioned in Sec.~\ref{sec:momentum}). 

At the receiving end of the cosmic cascade, the geometry and the response function of the detector itself need to be simulated, at different levels of detail depending on the use case: it is appropriate to approximate an entire muon telescope as a point in space, when the target to be imaged and its distance are orders of magnitude larger than the detector, while this is obviously not true for small-target applications. Similarly, simple parametrizations of the detector efficiency as a function of a few variables are usually sufficient, but high-precision use cases may require full simulations of the detector response (e.g. with GEANT4~\cite{GEANT4}) as customary in particle and nuclear physics. 

More importantly than what happens at the start or at the end of the muon trajectory, the simulation step that is most relevant to muography is the particle propagation through the target material. Multipurpose tools developed for particle and nuclear physics, such as GEANT4~\cite{GEANT4} or FLUKA~\cite{FLUKA}, allow a very accurate simulation of the muon interaction with the traversed material, but take a very long computation time at each simulation step. 
For this reason, parametric or semi-parametric simulations have been developed with the aim to reduce CPU time while preserving good accuracy. 
As a representative example, we elaborate on the specialized code MUSIC (MUon SImulation Code)~\cite{Antonioli1997,Kudryavtsev2009}, in use by several muography teams. MUSIC was developed for the use case of muon transport through large volumes of material and initially motivated by the study of muons as background to underground, underwater and under-ice neutrino experiments. MUSIC only considers the main electromagnetic interactions causing energy loss, namely ionisation including knock-on electron production, bremsstrahlung (in Born approximation), electron-positron pair production and photonuclear interactions (i.e. muon-nucleus inelastic scattering), while for example Coulomb corrections to bremsstrahlung are neglected (they do not exceed 1\% for heavy nuclei). All interaction processes are stochastically considered if the fraction of energy ($v$) lost by a muon in the interaction exceeds a user-defined threshold $v_{cut}$, while a continuous approximation is used below $v_{cut}$. This threshold must be tuned to reach a trade-off between speed and accuracy, and an optimal threshold is found to be of order $10^{-3}$~\cite{Kudryavtsev2009}. The program evaluates the mean free path of a muon between two subsequent interactions, then it samples the real path of the muon to the next interaction using a random number generator. Ionization is parametrized by the Bethe-Bloch formula (Eq.~\ref{eq:dedx}) for $v < v_{cut}$, while knock-on electron production is only considered for $v \ge v_{cut}$, as well as multiple Coulomb scattering that is treated in the Gaussian approximation (Eq.~\ref{eq:scattering}). 
All muons are considered to be ballistic, and if the total muon energy becomes smaller than $m_{\mu}c^2$, the muon is considered as stopped. 
An alternative version of MUSIC is dedicated to muon transport through thin slabs of materials, by treating all interactions as stochastic. This implies a much larger CPU usage, and yields more accurate predictions for momentum spectra and angular deviations in thin slabs of material while reproducing the results of the standard version for large thicknesses. 
The MUSUN (MUon Simulations UNderground)~\cite{Kudryavtsev2009} code convolutes the simulation of muon transport from MUSIC with simulations of the muon energy spectrum and angular distribution.

All the Monte Carlo approaches discussed so far in this Section execute each propagation step of the muon in a chronological order (``forward'' Monte Carlo), this although natural implies a very large degree of inefficiency, especially for large targets. In fact, in order to accurately simulate the flux of muons that are able to reach the detector (and in particular the low-momentum muons that can be in the acceptance after large-angle scattering, see Section~\ref{sec:bkg}) the program has to sample muons produced from all directions in the atmosphere, although the useful statistical sample comes only from the tiny fraction that passes through the O($1~{\rm m^2}$) surface of the detector.\newline
This problem is addressed by \virgol{backward Monte Carlo} programs, such as PUMAS (backward acronym for Semi-Analytical MUons Propagation)~\cite{PUMAS}, designed having in mind volcano muography use cases. 
PUMAS allows exclusive sampling of a final state by reversing the simulation flow. 
At each simulation step through the target material, the muon interactions with matter are split into a continuous component describing collective processes, such as multiple scattering and continuous energy loss, and one or more discrete interactions; the optimal threshold for transition between the two regimes depends on the application. The authors highlight a few case studies~\cite{PUMAS} where the outcomes of PUMAS, run in forward and backward mode, agree to better than 1\%, stating that PUMAS can achieve an accuracy comparable to GEANT4 but with a speedup of two orders of magnitude in backward mode.


%% file: conclusions.tex
\section{Conclusions and Outlook}
\label{sec:conclusions}

We reviewed the fundamental aspects and the state of the art of muography, showing how concepts that have their origin in particle physics are finding applications in the study of natural or man-made structures. 
Recent years have seen a rapid growth in the number of academic papers on the subject, as well as patents and commercialization attempts~\cite{Kaiser2018}, as several detector and analysis breakthroughs now allow an effective transition from ``proof of principle'' to mature applications. 

In this kind of research, there is always a risk of falling in the cognitive bias exemplified by the saying ``when you have a hammer, everything looks like a nail'':  
the fact that an object can be imaged by muography, does not always mean that this kind of imaging is useful. It is thus increasingly important, as the methods mature, to team up with members of the relevant user communities who can help state precisely the relevant research questions and steer any further development in the correct direction.

From the point of view of geosciences and archaeology, muography has to compete with more established remote-sensing methods, such as gravimetry, seismotomography, electrical resistivity, ground penetrating radar, etc. 
Nevertheless, muography is regarded by many geophysicists and archaeologists as a promising technique because of its complementary merits. In particular, its intrinsic directionality allows to form anyway an image of the target  even without relying on a conceptually complex and computationally intense non-linear inversion procedure. This can benefit standard geophysical methods, typically needing strong constraints and hence a significant degree of model dependence, in order to identify a unique solution to the inversion problem. 
Currently, volcanology is one of the most intensive areas under investigation, where many research questions related to different structures at different depths within the volcano can be addressed with muography, and where most teams are truly multidisciplinary, thus demonstrating the acceptance of this new method by the volcanological community. The imaging of glaciers recently entered the list of applications of muography, providing unique measurements of the inaccessible ice-bedrock interface. 

There is a practically inexhaustible number of targets of high geophysical or historical interest in the world, where imaging through muography might have a large potential impact. In practice, however, the priorities have often been influenced by local considerations, such as the proximity of the research team to a suitable target, or the presence of a strong community of potential users of the method in the same institution. 
Two trends can be observed in these areas, moving in opposite directions: the establishment of long-term monitoring by installing large-area detectors (e.g. for volcano surveillance), and the transition to a phase of commercialization for portable detectors, that might open a large variety of currently unexplored or under-explored applications of both societal and academic relevance.


Homeland security and nuclear safety are also very popular areas of investigation for muography (typically SM), with a direct societal impact. In general, muography has a clear advantage, in the detection of radioactive material placed in storage containers, over methods that rely on the detection of the radiation emitted by the material itself, because muons will pass through those containers that are opaque to the radiation they are designed to shield. Other targets of interests for SM range from industrial process control (e.g. blast furnaces) to structural integrity verification, where the penetrating power of muons allows to reach inaccessible structures. Once again, we are in the presence of a very lively field with some of the actors that have moved from research to university spin-off or industrial partnerships.

Finally we would like to point out that, while muography is nowadays evolving from the physicist laboratory to the  outside world of applications, there is still plenty of room for novel research.  
For sake of example, as atmospheric muons are almost all produced in the upper atmosphere, the middle atmosphere can be treated as the object of study, and phenomena happening within it~\cite{Grashorn:2009ey} may become the target of investigation~\cite{Hariharan:2019fai}.  
The development of compact and very robust portable muon detectors may even benefit the exploration of extra-terrestrial bodies~\cite{Kedar2013}. 
And many other areas may just be waiting for a clever application of this method.


%% file: acknowledgments.tex
We are grateful to Ciro Pistillo, S\'ebastien Procureur and Hiroyuki Tanaka who kindly pointed us to some of the references in our bibliography. 
Samip Basnet, Salvatore Giammanco and Sophie Wuyckens provided useful feedback on an early draft. 
A.G.'s work was partially supported by the EU Horizon 2020 Research and
Innovation Programme under the Marie Sklodowska-Curie Grant
Agreement No. 822185, and by the Fonds de la Recherche Scientifique - FNRS under Grant No. T.0099.19.

%% file: arxiv-template.bbl
\begin{thebibliography}{100}

\bibitem{Galison1983}
P.~Galison.
\newblock The discovery of the muon and the failed revolution against quantum
  electrodynamics.
\newblock {\em Centaurus}, 26:262, 1983.

\bibitem{Anderson1937}
C.~D. Anderson and S.~Neddermeyer.
\newblock Note on the nature of cosmic-ray particles.
\newblock {\em Phys. Rev. 51, 884}, 51:884, 1937.

\bibitem{Anderson1961}
C.~D. Anderson.
\newblock Early work on the positron and muon.
\newblock {\em Am. J. Phys.}, 29:825, 1961.

\bibitem{ConversiPanciniPiccioni}
E.~Conversi, M.~Pancini and O.~Piccioni.
\newblock On the disintegration of negative mesons.
\newblock {\em Phys. Rev.}, 71:209, 1947.

\bibitem{LattesMuirheadOcchialiniPowell}
C.~M.~G. Lattes, H.~Muirhead, G.~P.~S. Occhialini, and C.~F. Powell.
\newblock Processes involving charged mesons.
\newblock {\em Nature}, 159:694, 1947.

\bibitem{PDG2018}
M.~Tanabashi et~al.
\newblock Review of particle physics.
\newblock {\em Phys. Rev. D}, 98:030001, Aug 2018.

\bibitem{Feruglio:2015jfa}
Ferruccio Feruglio.
\newblock {Pieces of the Flavour Puzzle}.
\newblock {\em Eur. Phys. J. C}, 75(8):373, 2015.

\bibitem{George1955}
E.~P. George.
\newblock Cosmic rays measure overburden of tunnel.
\newblock {\em Commonwealth Engineer}, July 1:455, 1955.

\bibitem{Alvarez1970}
Luis~W. Alvarez et~al.
\newblock Search for hidden chambers in the pyramids.
\newblock {\em Science}, 167:832, 1970.

\bibitem{Malmqvist1979}
L.~Malmqvist, G.~J\"{o}nsson, K.~Kristiansson, and L.~Jacobsson.
\newblock Theoretical studies of in-situ rock density determination using
  cosmic-ray muon intensity measurements with application in mining geophysics.
\newblock {\em Geophys.}, 44(9):1549, 1979.

\bibitem{Nagamine1995}
K.~Nagamine, M.~Iwasaki, K.~Shimomura, and K.~Ishida.
\newblock Method of probing inner-structure of geophysical substance with the
  horizontal cosmic-ray muons and possible application to volcanic eruption
  prediction.
\newblock {\em Nucl. Inst. Meth. A}, 356:585, 1995.

\bibitem{Tanaka2009}
Hiroyuki Tanaka et~al.
\newblock Detecting a mass change inside a volcano by cosmic-ray muon
  radiography (muography): First results from measurements at asama volcano,
  japan.
\newblock {\em Geophys. Res.Lett.}, 36:1944, 2009.

\bibitem{LosAlamos2003}
Konstantin~N. Borozdin et~al.
\newblock Radiographic imaging with cosmic-ray muons.
\newblock {\em Nature}, 422:277, 2003.

\bibitem{Kaiser2018}
Ralf Kaiser.
\newblock Muography: overview and future directions.
\newblock {\em Phil. Trans. R. Soc. A}, 377:0049, 2018.

\bibitem{Procureur2018}
Sebastien Procureur.
\newblock Muon imaging: Principles, technologies and applications.
\newblock {\em Nucl. Inst. Meth. A}, 878:169, 2018.

\bibitem{CarloganuSaracino2012}
Giulio Saracino and Cristina Carloganu.
\newblock Looking at volcanoes with cosmic-ray muons.
\newblock {\em Physics Today}, 65:60, 2012.

\bibitem{Tanaka2017}
Hiroyuki Tanaka.
\newblock {Muography}.
\newblock {\em PoS}, KMI2017:026, 2017.

\bibitem{Checchia2016}
P.~Checchia.
\newblock Review of possible applications of cosmic muon tomography.
\newblock {\em JINST}, 11(12):C12072, 2016.

\bibitem{Bonechi2005}
L.~Bonechi et~al.
\newblock {Development of the ADAMO detector: test with cosmic rays at
  different zenith angles}.
\newblock In {\em {Proceedings, 29th International Cosmic Ray Conference (ICRC
  2005): Pune, India}}, pages 283--286, 2005.

\bibitem{Bogdanova2006}
L.~N. Bogdanova, M.~G. Gavrilov, V.~N. Kornoukhov, and A.~S. Starostin.
\newblock Cosmic muon flux at shallow depths underground.
\newblock {\em Phys. Atom. Nuclei}, 69(8):1293, 2006.

\bibitem{SatoKinWatanabe2017}
H.~{Sato}, T.~{Kin}, and Y.~{Watanabe}.
\newblock Investigation of environmental cosmic-ray muon spectrum in low energy
  region.
\newblock In {\em 2017 IEEE Nuclear Science Symposium and Medical Imaging
  Conference (NSS/MIC)}, page~1, 2017.

\bibitem{Olah:2017zgo}
Laszlo Ol\'ah, Hiroyuki Tanaka, and Dezso Varga.
\newblock {Investigation of background sources of muography}.
\newblock {\em PoS}, ICRC2017:347, 2018.

\bibitem{Lin2010}
Jeng-Wei Lin, Yen-Fu Chen, Rong-Jiun Sheu, and Shiang-Huei Jiang.
\newblock {Measurement of angular distribution of cosmic-ray muon fluence
  rate}.
\newblock {\em Nucl. Inst. Meth. A}, 619:24, 2010.

\bibitem{Grieder2001}
Peter K.~F. Grieder.
\newblock {\em Cosmic Rays at Earth -- Researcher's Reference Manual and Data
  Book}.
\newblock Elsevier, 2001.

\bibitem{LosAlamos2014}
Christopher Morris et~al.
\newblock Horizontal cosmic ray muon radiography for imaging nuclear threats.
\newblock {\em Nucl. Inst. Meth. B}, 330:42, 2014.

\bibitem{Rutherford1911}
E.~Rutherford.
\newblock The scattering of $\alpha$ and $\beta$ particles by matter and the
  structure of the atom.
\newblock {\em Lond. Edinb. Dubl. Phil. Mag.}, 21(125):669, 1911.

\bibitem{LynchDahl1991}
Gerald~R. Lynch and Orin~I. Dahl.
\newblock {Approximations to multiple Coulomb scattering}.
\newblock {\em Nucl. Inst. Meth. B}, 58:6, 1991.

\bibitem{Tsai1974}
Yung-Su Tsai.
\newblock Pair production and bremsstrahlung of charged leptons.
\newblock {\em Rev. Mod. Phys.}, 46:815, 1974.

\bibitem{GlueX}
Z.~Papandreou, R.~Hakobyan, and N.~Kolev.
\newblock {BCAL} radiation length calculations.
\newblock Technical Report 439, GlueX, 2005.

\bibitem{Zenoni:2014kva}
Aldo Zenoni.
\newblock {Historical building stability monitoring by means of a cosmic ray
  tracking system}.
\newblock In {\em {Proceedings, 4th International Conference on Advancements in
  Nuclear Instrumentation Measurement Methods and their Applications (ANIMMA
  2015): Lisbon, Portugal, April 20-24}}, page 7465542, 2015.

\bibitem{Tanaka2018}
H.~Tanaka.
\newblock {Japanese volcanoes visualized with muography}.
\newblock {\em Phil. Trans. R. Soc. A}, 377:2137, 2018.

\bibitem{Raffaello2018}
R.~D'Alessandro et~al.
\newblock {Volcanoes in Italy and the role of muon radiography}.
\newblock {\em Phil. Trans. R. Soc. A}, 377:0050, 2018.

\bibitem{Marteau:2015pxa}
Jacques Marteau et~al.
\newblock {Muon tomography applied to active volcanoes}.
\newblock {\em PoS}, PhotoDet2015:004, 2016.

\bibitem{Tioukov2019}
V.~Tioukov et~al.
\newblock {First muography of Stromboli volcano}.
\newblock {\em Sci. Rep.}, 9:6695, 2019.

\bibitem{Barberi2013}
F.~Barberi.
\newblock {Risk assessment of Vesuvius volcano}.
\newblock GIFT2013, General Assembly of the European Geosciences Union, Vienna
  (Austria),
  \url{http://static2.egu.eu/media/filer_public/2013/06/20/barberi.pdf}, 2013.

\bibitem{MURAY2014}
F.~Ambrosino et~al.
\newblock {The MU-RAY project: detector technology and first data from Mt.
  Vesuvius}.
\newblock {\em JINST}, 9:C02029, 2014.

\bibitem{Saracino:2017mao}
Giulio Saracino et~al.
\newblock {The MURAVES muon telescope: technology and expected performances}.
\newblock {\em Ann. Geophys. Italy}, 60(1):S0103, 2017.

\bibitem{Carloganu:2011zz}
Cristina C{\^a}rloganu.
\newblock {Density imaging of volcanoes with atmospheric muons using GRPCs}.
\newblock {\em PoS}, EPS-HEP2011:055, 2011.

\bibitem{Bene:2013fwa}
Samuel B\'en\'e et~al.
\newblock {Volcano radiography with GRPCs}.
\newblock In {\em {Proceedings, International Conference on Calorimetry for the
  High Energy Frontier (CHEF 2013): Paris, France, April 22-25, 2013}}, page
  414, 2013.

\bibitem{Carloganu2013}
C.~C{\^a}rloganu et~al.
\newblock {Towards a muon radiography of the Puy de D{\^o}me}.
\newblock {\em {Geosci. Inst. Meth. Data Syst.}}, 2:55, 2013.

\bibitem{Ambrosino:2015yqk}
F.~Ambrosino et~al.
\newblock {Joint measurement of the atmospheric muon flux through the Puy de
  D{\^o}me volcano with plastic scintillators and Resistive Plate Chambers
  detectors}.
\newblock {\em J. Geophys. Res. Solid Earth}, 120(11):7290, 2015.

\bibitem{JapanMeteorologicalAgency}
{Minutes of the 113th Volcanic Eruption Liaison Committee (in Japanese)}.
\newblock
  \url{https://www.data.jma.go.jp/svd/vois/data/tokyo/STOCK/kaisetsu/CCPVE/Report/103/kaiho_103_32.pdf},
  2009.
\newblock Date and time: 13:00-17:30 on June 18, 2009.

\bibitem{Jourde2016}
Kevin Jourde et~al.
\newblock {Muon dynamic radiography of density changes induced by hydrothermal
  activity at the La Soufri\`ere of Guadeloupe volcano}.
\newblock {\em Sci. Rep.}, 6:33406, 2016.

\bibitem{Marteau:2016jcn}
Jacques Marteau et~al.
\newblock {DIAPHANE: Muon tomography applied to volcanoes, civil engineering,
  archaelogy}.
\newblock {\em JINST}, 12(02):C02008, 2017.

\bibitem{Gonidec:2018kij}
Y.~Le~Gonidec et~al.
\newblock {Abrupt changes of hydrothermal activity in a lava dome detected by
  combined seismic and muon monitoring}.
\newblock {\em Sci. Rep.}, 9(1):3079, 2019.

\bibitem{Carbone2013Etna}
Daniele Carbone et~al.
\newblock {An experiment of muon radiography at Mt Etna (Italy)}.
\newblock {\em Geophys. J. Int.}, 196(2):633, 10 2013.

\bibitem{MEV2018}
D.~{Lo Presti} et~al.
\newblock The {MEV} project: Design and testing of a new high-resolution
  telescope for muography of {Etna} volcano.
\newblock {\em Nucl. Inst. Meth. A}, 904:195, 2018.

\bibitem{Catalano:2015zxd}
Osvaldo Catalano et~al.
\newblock {Volcanoes muon imaging using Cherenkov telescopes}.
\newblock {\em Nucl. Inst. Meth. A}, 807:5, 2016.

\bibitem{DelSanto:2017ytw}
M.~Del~Santo et~al.
\newblock {Looking inside volcanoes with the Imaging Atmospheric Cherenkov
  Telescopes}.
\newblock {\em Nucl. Inst. Meth. A}, 876:111, 2017.

\bibitem{CTA}
{CTA Consortium}.
\newblock {Design concepts for the Cherenkov Telescope Array CTA: An advanced
  facility for ground-based high-energy gamma-ray astronomy}.
\newblock {\em Exp. Astron.}, 32(3):193, 2011.

\bibitem{Azuma2014}
K.~Azuma, H.~Tanaka, H.~Suenaga, and K.~Suzuki.
\newblock Muographic test measurements for monitoring groundwater.
\newblock In {\em {ISRM International Symposium - 8th Asian Rock Mechanics
  Symposium, 14-16 October, Sapporo, Japan}}, 2014.

\bibitem{Tanaka2011}
Hiroyuki Tanaka et~al.
\newblock {Cosmic muon imaging of hidden seismic fault zones: Rainwater
  permeation into the mechanical fractured zones in Itoigawa-Shizuoka Tectonic
  Line, Japan}.
\newblock {\em Earth Planet. Sci. Lett.}, 306(3):156, 2011.

\bibitem{Lesparre2016}
N.~Lesparre et~al.
\newblock Electrical resistivity imaging in transmission between surface and
  underground tunnel for fault characterization.
\newblock {\em J. Appl. Geophys.}, 128:163, 2016.

\bibitem{Hivert2017}
Fanny Hivert et~al.
\newblock Muography sensitivity to hydrogeological rock density perturbation:
  roles of the absorption and scattering on the muon flux measurement
  reliability.
\newblock {\em Near Surf. Geophys.}, 15(2):121, 2017.

\bibitem{Pazzi_EGU2019}
V.~Pazzi.
\newblock {Oral presentation at the EGU General Assembly 2019}.
\newblock
  \url{https://meetingorganizer.copernicus.org/EGU2019/EGU2019-8784.pdf}, 2019.

\bibitem{Ihl:2010uv}
Matthias Ihl.
\newblock {Muon Tomography of Ice-filled Cleft Systems in Steep Bedrock
  Permafrost: A Proposal}, 2010.
\newblock {arXiv:1008.1241 [physics.geo-ph]}.

\bibitem{Klinger:2015gva}
J.~Klinger et~al.
\newblock {Simulation of muon radiography for monitoring CO$_2$ stored in a
  geological reservoir}.
\newblock {\em Int. J. Greenhouse Gas Control}, 42:644, 2015.

\bibitem{Gluyas2018}
J.~Gluyas et~al.
\newblock {Passive, continuous monitoring of carbon dioxide geostorage using
  muon tomography}.
\newblock {\em Phil. Trans. R. Soc. A}, 377:0059, 2018.

\bibitem{Caffau1997}
E.~Caffau, F.~Coren, and G.~Giannini.
\newblock {Underground cosmic-ray measurement for morphological reconstruction
  of the “Grotta Gigante” natural cave}.
\newblock {\em Nucl. Inst. Meth. A}, 385(3):480, 1997.

\bibitem{Olah2012}
L.~Ol\'ah et~al.
\newblock {CCC-based muon telescope for examination of natural caves}.
\newblock {\em Geosci. Inst. Meth. Data Syst.}, 1(2):229, 2012.

\bibitem{Schouten2018}
Doug Schouten.
\newblock Muon geotomography: selected case studies.
\newblock {\em Phil. Trans. R. Soc. A}, 377:0061, 2018.

\bibitem{Lingacom2018}
A.~Harel and D.~Yaish.
\newblock Lingacom muography.
\newblock {\em Phil. Trans. R. Soc. A}, 377:0133, 2018.

\bibitem{Nishiyama2017}
R.~Nishiyama et~al.
\newblock First measurement of ice-bedrock interface of alpine glaciers by
  cosmic muon radiography.
\newblock {\em Geophys. Res. Lett.}, 44(12):6244, 2017.

\bibitem{Nishiyama2019}
R.~Nishiyama et~al.
\newblock Bedrock sculpting under an active alpine glacier revealed from
  cosmic-ray muon radiography.
\newblock {\em Sci. Rep.}, 9(1):6970, 2019.

\bibitem{Ariga2018}
Akitaka Ariga et~al.
\newblock A nuclear emulsion detector for the muon radiography of a glacier
  structure.
\newblock {\em Instruments}, 2(2), 2018.

\bibitem{Huss2013}
M.~Huss.
\newblock Density assumptions for converting geodetic glacier volume change to
  mass change.
\newblock {\em Cryosphere}, 7(3):877, 2013.

\bibitem{Gomez:2016kye}
H.~G\'omez et~al.
\newblock {Studies on muon tomography for archaeological internal structures
  scanning}.
\newblock {\em J. Phys. Conf. Ser.}, 718(5):052016, 2016.

\bibitem{Kyushu2018}
Kullapha Chaiwongkhot et~al.
\newblock {Development of a Portable Muography Detector for Infrastructure
  Degradation Investigation}.
\newblock {\em IEEE Trans. Nucl. Sci.}, 65:2316, 2018.

\bibitem{Nagamine2005-furnaces}
Kanetada Nagamine et~al.
\newblock Probing the inner structure of blast furnaces by cosmic-ray muon
  radiography.
\newblock {\em Proc. Japan Acad. B}, 81(7):257, 2005.

\bibitem{Vanini2018}
S.~Vanini et~al.
\newblock Muography of different structures using muon scattering and
  absorption algorithms.
\newblock {\em Phil. Trans. R. Soc. A}, 377:0051, 2018.

\bibitem{Guardincerri:2016mrk}
E.~Guardincerri et~al.
\newblock {Imaging the inside of thick structures using cosmic rays}.
\newblock {\em AIP Adv.}, 6:015213, 2016.

\bibitem{MuonSystems2018}
Pablo {Martinez Ruiz-del Arbol}, Pablo {Gomez Garcia}, Carlos {Diez Gonzalez},
  and Aitor {Orio Alonso}.
\newblock Non-destructive testing of industrial equipment using muon
  radiography.
\newblock {\em Phil. Trans. R. Soc. A}, 377:0054, 2018.

\bibitem{Menchaca-Rocha:2014yxa}
Arturo Menchaca-Rocha.
\newblock {Using cosmic muons to search for cavities in the Pyramid of the Sun,
  Teotihuacan: preliminary results}.
\newblock In {\em {Proceedings, 10th Latin American Symposium on Nuclear
  Physics and Applications: Montevideo, Uruguay, December 1-6, 2013}}, volume
  XLASNPA, page 012, 2014.

\bibitem{Melesio2014}
Lucina Melesio.
\newblock The pyramid detectives.
\newblock {\em Phys. World}, 27(12):24, dec 2014.

\bibitem{MayaMuonBook}
Roy Schwitters et~al.
\newblock {\em The {University} of {Texas} {Maya} {Muon} {Project}}.
\newblock Fermi National Laboratory, United States Department of Energy, Office
  of Science, 2007.

\bibitem{MayaMuonWeb}
{homepage}.
\newblock \url{http://www.hep.utexas.edu/mayamuon/}, Accessed: May 2019.

\bibitem{ScanPyramids}
{English homepage}.
\newblock \url{http://www.scanpyramids.org/index-en.html}, Accessed: May 2019.

\bibitem{Morishima:2017ghw}
Kunihiro Morishima et~al.
\newblock {Discovery of a big void in Khufu's Pyramid by observation of
  cosmic-ray muons}.
\newblock {\em Nature}, 552(7685):386, 2017.

\bibitem{Saracino2017-Echia}
G.~Saracino et~al.
\newblock Imaging of underground cavities with cosmic-ray muons from
  observations at {Mt}. {Echia} ({Naples}).
\newblock {\em Sci. Rep.}, 7(1):1181, April 2017.

\bibitem{Cimmino2019}
Luigi Cimmino et~al.
\newblock {3D Muography for the Search of Hidden Cavities}.
\newblock {\em Sci. Rep.}, 9:2974, 2019.

\bibitem{MURAY2013}
A.~Anastasio et~al.
\newblock {The MU-RAY experiment. An application of SiPM technology to the
  understanding of volcanic phenomena}.
\newblock {\em Nucl. Inst. Meth. A}, 718:134, 2013.

\bibitem{Baccani:2018nrn}
Guglielmo Baccani et~al.
\newblock {The MIMA project. Design, construction and performances of a compact
  hodoscope for muon radiography applications in the context of Archaeology and
  geophysical prospections}.
\newblock {\em JINST}, 13(11):P11001, 2018.

\bibitem{Riggi:2017izf}
F.~Riggi et~al.
\newblock {The Muon Portal Project: Commissioning of the full detector and
  first results}.
\newblock In {\em {8th International Conference on New Developments in
  Photodetection (NDIP17), Tours, France, July 3-7, 2017}}, volume 912,
  page~16, 2018.

\bibitem{Checchia2018}
P.~Checchia et~al.
\newblock {INFN} muon tomography demonstrator: past and recent results with an
  eye to near-future activities.
\newblock {\em Phil. Trans. R. Soc. A}, 377:0065, 2018.

\bibitem{Glasser:2018rhl}
V.~Glasser and R.~Lipton.
\newblock {Data analysis and detector troubleshooting for the Silicon Muon
  Scanner}.
\newblock Technical Report FERMILAB-PUB-18-503-E, FNAL, 2018.

\bibitem{DecisionSciences}
{Decision Sciences}.
\newblock {``About us'' page}.
\newblock \url{https://decisionsciences.com/about-us/}, Accessed: May 2019.

\bibitem{Poulson:2016fre}
D.~Poulson et~al.
\newblock {Cosmic ray muon computed tomography of spent nuclear fuel in dry
  storage casks}.
\newblock {\em Nucl. Inst. Meth. A}, 842:48, 2017.

\bibitem{Poulson2018}
D.~Poulson et~al.
\newblock {Application of muon tomography to fuel cask monitoring}.
\newblock {\em Phil. Trans. R. Soc. A}, 377:0052, 2018.

\bibitem{Ambrosino:2014aaa}
F.~Ambrosino et~al.
\newblock {Assessing the Feasibility of Interrogating Nuclear Waste Storage
  Silos using Cosmic-ray Muons}.
\newblock {\em JINST}, 10(06):T06005, 2015.

\bibitem{Clarkson2015}
A.~Clarkson et~al.
\newblock Characterising encapsulated nuclear waste using cosmic-ray muon
  tomography.
\newblock {\em JINST}, 10(03):P03020, mar 2015.

\bibitem{Mahon2018}
D.~Mahon et~al.
\newblock {First-of-a-kind muography for nuclear waste characterization}.
\newblock {\em Phil. Trans. R. Soc. A}, 377:0048, 2018.

\bibitem{Borozdin2012}
Konstantin Borozdin et~al.
\newblock {Cosmic Ray Radiography of the Damaged Cores of the Fukushima
  Reactors}.
\newblock {\em Phys. Rev. Lett.}, 109:152, 2012.

\bibitem{Miyadera2013}
Haruo Miyadera et~al.
\newblock {Imaging Fukushima Daiichi reactors with muons}.
\newblock {\em AIP Advances}, 3(5):052133, 2013.

\bibitem{Kume:2016exx}
N.~Kume et~al.
\newblock {Muon trackers for imaging a nuclear reactor}.
\newblock {\em JINST}, 11(09):P09008, 2016.

\bibitem{Fujii:2019kwi}
Hirofumi Fujii et~al.
\newblock {Imaging the Inner Structure of a Nuclear Reactor by Cosmic Muon
  Radiography}, 2019.
\newblock {arXiv:1902.01992 [physics.ins-det]}.

\bibitem{WNN2015}
{Muon data confirms fuel melt at Fukushima Daiichi 1}.
\newblock
  \url{http://www.world-nuclear-news.org/RS-Muon-data-confirms-fuel-melt-at-Fukushima-Daiichi-1-2303154.html},
  23 March 2015.

\bibitem{WNN2017}
{Muons suggest location of fuel in unit 3}.
\newblock
  \url{https://www.world-nuclear-news.org/RS-Muons-suggest-location-of-fuel-in-unit-3-0210174.html},
  02 October 2017.

\bibitem{Morishima-Fukushima}
Kunihiro Morishima.
\newblock {\em Muographic investigation in Fukushima nuclear power plant},
  volume Muography: Perspective Drawing in the 21st Century.
\newblock University of Tokyo Museum, 2015.
\newblock Edited by Hideaki Miyamoto, Takafumi Niihara, Hiroyuki Tanaka.

\bibitem{Hamawebinar}
S.~Piatek.
\newblock {Silicon Photomultiplier. Operation, Performance \& Possible
  Applications}.
\newblock \url{https://www.hamamatsu.com/sp/hc/osh/sipm_webinar_1.10.pdf},
  Accessed: May 2019.

\bibitem{Borshchev2017}
O.~Borshchev et~al.
\newblock Development of a new class of scintillating fibres with very short
  decay time and high light yield.
\newblock {\em JINST}, 12:P05013, 2017.

\bibitem{Bonneville2018}
A.~Bonneville et~al.
\newblock Borehole muography of subsurface reservoirs.
\newblock {\em Phil. Trans. R. Soc. A}, 377:0060, 2018.

\bibitem{Battiston2006}
S.~Ansoldi et~al.
\newblock {MGR: An innovative, low-cost and compact cosmic-ray detector}.
\newblock {\em Nucl. Inst. Meth. A}, 567(1):298, 2006.

\bibitem{bozzaC-nucl_emul}
Cristiano Bozza et~al.
\newblock Nuclear emulsion techniques for muography.
\newblock {\em Ann. Geophys.}, 60:1, 2017.

\bibitem{Charpak}
Georges Charpak et~al.
\newblock Investigation of some properties of multiwire proportional chambers.
\newblock {\em Nucl. Inst. Meth.}, 88(1):149, 1970.

\bibitem{OlahHamarMiyamotoTanaka2018}
Laszlo Ol\'ah et~al.
\newblock The first prototype of an {MWPC}-based borehole-detector and its
  application for muography of an underground pillar.
\newblock {\em BUTSURI-TANSA (Geophysical Exploration)}, 71:161, 2018.

\bibitem{Bouteille2016}
S.~Bouteille et~al.
\newblock {A Micromegas-based telescope for muon tomography: The WatTo
  experiment}.
\newblock {\em Nucl. Inst. Meth. A}, 834:223, 2016.

\bibitem{LazaroRoche2016-micromegas}
I.~{L\'azaro Roche} et~al.
\newblock {Overview and outlook on muon survey tomography based on Micromegas
  detectors for unreachable sites technology}.
\newblock {\em E3S Web of Conf.}, 12:03001, 2016.

\bibitem{Procureur:2013yea}
S.~Procureur, R.~Dupr\'e, and S.~Aune.
\newblock {Genetic multiplexing and first results with a $50$x$50~{\rm cm}^2$
  Micromegas}.
\newblock {\em Nucl. Inst. Meth. A}, 729:888, 2013.

\bibitem{Bouteille:2016wdv}
S.~Bouteille et~al.
\newblock {Large resistive 2D Micromegas with genetic multiplexing and some
  imaging applications}.
\newblock {\em Nucl. Inst. Meth. A}, 834:187, 2016.

\bibitem{Hivert2015}
F.~Hivert et~al.
\newblock {Temporal tomography of rock density using muon measurements with
  TPC-Micromegas}.
\newblock In {\em {ISRM}-13CONGRESS-2015-212}, page~9, ISRM, January 2015.
  International Society for Rock Mechanics and Rock Engineering.

\bibitem{Baesso}
P.~Baesso et~al.
\newblock A high resolution resistive plate chamber tracking system developed
  for cosmic ray muon tomography.
\newblock {\em JINST}, 8:P08006, 2013.

\bibitem{Wuyckens2018}
S.~Wuyckens, A.~Giammanco, P.~Demin, and E.~{Cortina Gil}.
\newblock {A portable muon telescope based on small and gas-tight Resistive
  Plate Chambers}.
\newblock {\em Phil. Trans. R. Soc. A}, 377:0139, 2018.

\bibitem{Kedar2013}
S.~Kedar et~al.
\newblock {Muon radiography for exploration of Mars geology}.
\newblock {\em Geosci. Inst. Meth. Data Syst.}, 2(1):157, 2013.

\bibitem{Temperino2019}
Guglielmo Baccani et~al.
\newblock {Muon Radiography of Ancient Mines: The San Silvestro Archaeo-Mining
  Park (Campiglia Marittima, Tuscany)}.
\newblock In {\em {Selected Papers from the 7th International Conference on New
  Frontiers in Physics (ICNFP 2018)}}, volume~5, 2019.

\bibitem{Nishiyama2016}
Ryuichi Nishiyama, Akimichi Taketa, Seigo Miyamoto, and Katsuaki Kasahara.
\newblock {Monte Carlo simulation for background study of geophysical
  inspection with cosmic-ray muons}.
\newblock {\em Geophys. J. Int.}, 206(2):1039, 2016.

\bibitem{Tarantola1982}
A.~Tarantola et~al.
\newblock {Generalized nonlinear inverse problems solved using the
  least-squares criterion}.
\newblock {\em Rev. Geophys. Space Phys.}, 20(2):219, 1982.

\bibitem{Press1992}
W.H. Press et~al.
\newblock {\em {Numerical recipes in C: 3rd Edition}}.
\newblock 2007.

\bibitem{Tarantola2005}
A.~Tarantola.
\newblock {\em {Inverse Problem Theory and Methods for Model Parameter
  Estimation}}.
\newblock 2005.

\bibitem{Tanaka2010}
H.~Tanaka et~al.
\newblock {Three-dimensional computational axial tomography scan of a volcano
  with cosmic ray muon radiography}.
\newblock {\em J. Geophys. Res.}, 115(B12332), 201.

\bibitem{Nishiyama2014}
R.~Nishiyama et~al.
\newblock {Integrated processing of muon radiography and gravity anomaly data
  toward the realization of high-resolution 3-D density structural analysis of
  volcanoes: Case study of Showa-Shinzan lava dome, Usu, Japan}.
\newblock {\em J. Geophys. Res.: Solid Earth}, 119(1):699, 2014.

\bibitem{Tanaka2015}
H.~Tanaka et~al.
\newblock {Muographic mapping of the subsurface density structures in Miura,
  Boso and Izu peninsulas, Japan}.
\newblock {\em Sci. Rep.}, 5:8305, 2015.

\bibitem{Lesparre2012}
N.~Lesparre et~al.
\newblock {Density muon radiography of La Soufri{\`e}re of Guadeloupe volcano:
  comparison with geological, electrical resistivity and gravity data}.
\newblock {\em Geophys. J. Int.}, 190:1008, 2012.

\bibitem{Bryman2014}
Douglas Bryman et~al.
\newblock {Muon Geotomography - Bringing New Physics to Orebody Imaging}.
\newblock In {\em {Building Exploration Capability for the 21st Century}}.
  Society of Economic Geologists, 01 2014.

\bibitem{Guardincerri2017}
Elena Guardincerri et~al.
\newblock {3D Cosmic Ray Muon Tomography from an Underground Tunnel}.
\newblock {\em Pure and Appl. Geophys.}, 174:2133, 2017.

\bibitem{Bonechi2015back_projections}
L.~Bonechi et~al.
\newblock A projective reconstruction method of underground or hidden
  structures using atmospheric muon absorption data.
\newblock {\em JINST}, 10:P02003, 2015.

\bibitem{Bonechi2018UK}
L.~Bonechi et~al.
\newblock Tests of a novel imaging algorithm to localize hidden objects or
  cavities with muon radiography.
\newblock {\em Phil. Trans. R. Soc. A}, 377:0063, 2018.

\bibitem{Morris2003}
C.~Morris et~al.
\newblock {Detection of high-$Z$ objects using multiple scattering of cosmic
  ray muons}.
\newblock {\em Rev. Sci. Inst.}, 74:4294, 2003.

\bibitem{Chatzidas2016}
S.~Chatzidakis et~al.
\newblock {Analysis of spent nuclear fuel imaging using multiple Coulomb
  scattering of cosmic muons}.
\newblock {\em IEEE Trans. Nucl. Sci.}, 63:6, 2016.

\bibitem{Rosas-Carbajal:2017sdu}
Marina Rosas-Carbajal et~al.
\newblock {Three-dimensional density structure of La Soufri{\`e}re de
  Guadeloupe lava dome from simultaneous muon radiographies and gravity data}.
\newblock {\em Geophys. Res. Lett.}, 44(13):6743, 2017.

\bibitem{Lelievre2019-joint}
Peter~G Leli\`evre et~al.
\newblock {Joint inversion methods with relative density offset correction for
  muon tomography and gravity data, with application to volcano imaging}.
\newblock {\em Geophys. J. Int.}, 2019.

\bibitem{Cella2007}
Federico Cella et~al.
\newblock {Shallow structure of the Somma-Vesuvius volcano from 3D inversion of
  gravity data}.
\newblock {\em J. Volcanol. Geotherm. Res.}, 161(4):303, 2007.

\bibitem{Denatale2006}
G.~De~Natale et~al.
\newblock {The Somma-Vesuvius volcano (Southern Italy): Structure, dynamics and
  hazard evaluation}.
\newblock {\em Earth Sci. Rev.}, 74(1-2):73, 2006.

\bibitem{Calvari2014}
S.~Calvari et~al.
\newblock {Major eruptive style changes induced by structural modifications of
  a shallow conduit system: The 2007-2012 Stromboli case}.
\newblock {\em Bull. Volcanology}, 76:841, 2014.

\bibitem{Olah2017}
L\'aszl\'o Ol\'ah and Dezso Varga.
\newblock Investigation of soft component in cosmic ray detection.
\newblock {\em Astropart. Phys.}, 93:17, 2017.

\bibitem{Gomez:2017omj}
H.~G\'omez et~al.
\newblock {Forward scattering effects on muon imaging}.
\newblock {\em JINST}, 12(12):P12018, 2017.

\bibitem{Carbone2013}
K.~Jourde et~al.
\newblock Experimental detection of upward going cosmic particles and
  consequences for correction of density radiography of volcanoes.
\newblock {\em Geophys. Res. Lett.}, 40(24):6334, 2013.

\bibitem{Marteau:2013qua}
J.~Marteau et~al.
\newblock {Implementation of sub-nanosecond time-to-digital convertor in
  field-programmable gate array: applications to time-of-flight analysis in
  muon radiography}.
\newblock {\em Measur. Sci. Tech.}, 25:035101, 2014.

\bibitem{Cecchini:1998kv}
S.~Cecchini and M.~Sioli.
\newblock {Cosmic ray muon physics}.
\newblock In {\em {Non-accelerator particle astrophysics. Proceedings, 5th ICTP
  School, Trieste, Italy, June 29-July 10}}, page 201, 1998.

\bibitem{Hebbeker2002}
Thomas Hebbeker and Charles Timmermans.
\newblock A compilation of high energy atmospheric muon data at sea level.
\newblock {\em Astropart. Phys.}, 18(1):107, 2002.

\bibitem{Hariharan:2019fai}
B.~Hariharan et~al.
\newblock {Measurement of the Electrical Properties of a Thundercloud Through
  Muon Imaging by the GRAPES-3 Experiment}.
\newblock {\em Phys. Rev. Lett.}, 122:105101, 2019.

\bibitem{Adamson:2009zf}
P.~Adamson et~al.
\newblock {Observation of muon intensity variations by season with the MINOS
  far detector}.
\newblock {\em Phys. Rev.}, D81:012001, 2010.

\bibitem{Grashorn:2009ey}
E.~W. Grashorn et~al.
\newblock {The atmospheric charged kaon/pion ratio using seasonal variation
  methods}.
\newblock {\em Astropart. Phys.}, 33:140, 2010.

\bibitem{Bonechi:2003bi}
L.~Bonechi, M.~Grandi, F.~Taccetti, and E.~Vannuccini.
\newblock {ADAMO, an altazimuthal detector for atmospheric cosmic-ray
  observation}.
\newblock In {\em {Proceedings, 28th International Cosmic Ray Conference (ICRC
  2003): Tsukuba, Japan, 2003}}, pages 3485--3488, 2003.

\bibitem{NEWCUT}
{NEWCUT Hungary-Japan Laboratory Opens}.
\newblock
  \url{https://news.muographix.u-tokyo.ac.jp/2018/12/13/muographers-2018-newcut-hungary-japan-laboratory-official-opening-ceremony/},
  Accessed: May 2019.

\bibitem{CORSIKA}
D.~Heck et~al.
\newblock {CORSIKA: A Monte Carlo Code to Simulate Extensive Air Showers}.
\newblock Technical Report FZKA 6019, Forschungszentrum Karlsruhe, 1998.

\bibitem{GEANT4}
S.~Agostinelli et~al.
\newblock {GEANT4} --- a simulation toolkit.
\newblock {\em Nucl. Inst. Meth. A}, 506:250, 2003.

\bibitem{FLUKA}
T.T. B{\"o}hlen et~al.
\newblock {The FLUKA code: developments and challenges for high energy and
  medical applications}.
\newblock {\em Nucl. Data Sheets}, 120:211, 2014.

\bibitem{Antonioli1997}
P.~Antonioli et~al.
\newblock A three-dimensional code for muon propagation through the rock:
  {MUSIC}.
\newblock {\em Astropart. Phys.}, 7(4):357, 1997.

\bibitem{Kudryavtsev2009}
V.A. Kudryavtsev.
\newblock {Muon simulation codes MUSIC and MUSUN for underground physics}.
\newblock {\em Comput. Phys. Comm.}, 180(3):339, 2009.

\bibitem{PUMAS}
Valentin Niess, Anne Barnoud, Cristina Carloganu, and Eve Le~M\'en\'edeu.
\newblock {Backward Monte-Carlo applied to muon transport}.
\newblock {\em Comput. Phys. Comm.}, 229:54, 2018.

\end{thebibliography}
